\documentclass[aps,preprint,floats,epsf,epsfig,nofootinbib,letter]{revtex4}

\usepackage{graphicx}
\usepackage{dcolumn}
\usepackage{bm}
\usepackage{subfigure}
\usepackage{caption}
\usepackage{ragged2e}
\usepackage{hyperref}
\usepackage{amsmath}
\usepackage{mathrsfs}
\usepackage{color}

\def\be{\begin{eqnarray}}
\def\en{\end{eqnarray}}
\def\non{\nonumber}

\begin{document}

\renewcommand{\baselinestretch}{1.10}

\font\el=cmbx10 scaled \magstep2{\obeylines\hfill\today}

\vskip 1.5 cm

\centerline{\Large\bf Time-independent Green's Function} 
\centerline{\Large\bf of a Quantum Simple Harmonic Oscillator System and}
\centerline{\Large\bf Solutions with Additional Generic Delta-Function Potentials}

\bigskip
\centerline{\bf Chun-Khiang Chua, Yu-Tsai Liu, Gwo-Guang Wong}
\medskip
\centerline{Department of Physics and Chung Yuan Center for High Energy Physics,} 
\centerline{Chung Yuan Christian University,}
\centerline{Chung-Li, Taoyuan, Taiwan 32023, Republic of China}
\medskip

\centerline{\bf Abstract}
The one-dimensional time-independent Green's function $G_0$ of a quantum simple harmonic oscillator system ($V_0(x)=m \omega^2 x^2/2$) can be obtained by solving the equation directly. 
It has a compact expression, which gives correct eigenvalues and eigenfunctions easily. 
The Green's function $G$ with an additional delta-function potential can be obtained readily.
The same technics of solving the Green's function $G_0$ can be used to solve the eigenvalue problem of the simple harmonic oscillator with an generic delta-function potential at an arbitrary site, i.e. $V_1(x)\propto \delta(x-a)$. 
The Wronskians play an important and interesting role in the above studies. 
Furthermore, the approach can be easily generalized to solve the quantum system of a simple harmonic oscillator with two or more generic delta-function potentials.
We give the solutions of the case with two additional delta-functions for illustration.
 
\bigskip
\small

\pacs{Valid PACS appear here}

\maketitle

\section{Introduction}

The one-dimensional quantum simple harmonic oscillator (SHO) has become an indispensable material on the textbooks of quantum mechanics (for example, see~\cite{Gasiorowicz}) and widely used in many different physics and chemistry fields~\cite{AA,BB}. 
It is one of the most important model systems in quantum mechanics since any binding potential can usually be approximated as a harmonic potential at the vicinity of a stable equilibrium point.
It can be applied to the vibration of diatomic molecule, the Hooke's atom~\cite{KS}, the vibrations of atoms in a solid~\cite{Franz2008}, 
the quantum Hall effect~\cite{Laughlin1981}, the atoms in optical traps~\cite{Yannouleas1999}
and so on. 
The Schr\"odinger differential equation for a quantum SHO system can be analytically solved using either the Frobenius method~\cite{Lebedev} with an infinite series expansion or the algebra method~\cite{Griffiths} with the creation and annihilation operators to solve the eigenfunctions and the corresponding eigenvalues. 
Hence SHO is naturally to have research as well as pedagogical values.~\footnote{Sidney Coleman once said: ``The career of a young theoretical physicist consists of treating the harmonic oscillator in ever-increasing levels of abstraction."~\cite{Coleman}}

SHO is always a topic of interest. There are much to be investigated on this old, simple but important subject.
For example, it has been shown that the one-dimensional quantum harmonic oscillator problem is examined via the Laplace transform method. The stationary states are determined by requiring definite parity and good behaviour of the eigenfunction at the origin and at infinity \cite{Pimentel de Castro}.
Recently, in Ref.~\cite{Nogueira de Castro}, the exponential Fourier approach in the literature to the one-dimensional quantum harmonic oscillator problem is revised and criticized. The problem is revisited via the Fourier sine and cosine transform method and the stationary states are properly determined by requiring definite parity and square-integrable eigenfunctions \cite{Nogueira de Castro}.

Ref~\cite{Viana-Gomes} pointed out that an additional boundary condition neglected in the usual quantum mechanics textbooks should be imposed.   
It was shown that the following wave function satisfies the boundary condition and the Schr\"odinger equation of SHO ($V(x)=m \omega^2 x^2/2$) for $x\neq 0$,
\be 
u_\nu(x)=Ae^{-\frac{1}{2}(\alpha x)^2}U\left(-\frac{\nu}{2},\frac{1}{2},(\alpha x)^2\right),
\label{eq: realv}
\en
where $\alpha=\sqrt{m\omega/\hbar}$ and $U(a,b,x)$ is the Tricomi's (confluent hypergeometric) function. 
Requiring the derivative of the wave function be continuous at $x=0$, as implicated by the Schr\"odinger equation around $x=0$, gives $\nu=n=0,1,2,\cdots$ and
\be
u_\nu(x)=u_n(x)\propto e^{-\frac{1}{2}(\alpha x)^2}H_n(\alpha x).
\en
For non-integer $\nu$ the derivative of the wave function is discontinuous, as a byproduct it can be used to constructed the solution of the Schr\"odinger equation of SHO with a delta-function potential at the origin. 
In ref.~\cite{Patil} a harmonic oscillator with a $\delta(x)$ potential is analysed and compact expressions for the energy eigenvalues of the even parity states are also obtained.
However, these authors only considered the case of SHO with a delta-function potential located at the origin.

In this work, we use Green's function to solve the eigenvalue problem of a quantum SHO system 
in a new way, which is not given in any textbook and in the literature. 
For an introduction of Green's functions in quantum mechanics, one is referred to ref.~\cite{Economou,Sakurai}.
The important role of Wronskian played in obtaining the energy eigenvalues and eigenfunctions will be shown and hopefully be appreciated.  
Using the very same technics, one can easily solve the eigenvalue problem for a quantum SHO with one (two) additional delta-function potential(s) at an arbitrary site (arbitrary sites) directly from solving the Schr\"odinger equation.
In this way,  
we can have better understanding on the quantum SHO system with or without delta-function potentials. 
Some of our results can be found in~\cite{YTLiu}, which is prepared by one of the authors (YTL) based on a preliminary version of the present work. 

This paper is organized as follows. In Sec. II and III, we derive the time-independent Green's functions for the quantum systems of pure SHO, and SHO with a generic delta-function potential at an arbitrary site, respectively. From the poles of these Green's functions, we obtain the eigenvalues and eigenfunctions corresponding to the related quantum systems. In Sec IV and V, we derive the wave functions directly from the Schr\"odinger equation for the quantum systems of SHO with one and two generic delta-function(s), respectively. Sec. IV is the conclusion.

\section{Time-independent Green's function of a quantum simple harmonic oscillator}

The time-independent Green's function of a quantum simple harmonic oscillator satisfies the following equation:
\be
\left(-\frac{\hbar^2}{2m}\frac{\partial^2}{\partial x^2}+\frac{1}{2} m\omega^2 x^2-E \right)G_0(x,y; E)=\delta(x-y).
\en
By changing the variables, 
\be
\alpha\equiv \sqrt{\frac{m\omega}{\hbar}},
\quad
\varepsilon\equiv\frac{m}{\alpha^2\hbar^2} E=\frac{E}{\hbar\omega},
\label{eq: cvariables}
\en
the original equation can be expressed as
\be
\left(\frac{\partial^2}{\partial x^2}-\alpha^4 x^2+2\alpha^2\varepsilon \right)G_0(x,y; E)=-\frac{2m}{\hbar^2}\delta(x-y).
\label{eq: G0Eq}
\en
The Green's function needs to satisfy the following boundary conditions:
\be
\lim_{x\to\pm\infty} G_0(x,y; E)=0,
\en
and the matching conditions around $x=y$:
\be
G_0(x&=&y^+,y, E)=G_0(x=y^-,y;E),
\\
\Delta G'_0(y;E)\equiv
\frac{\partial}{\partial x} G_0(x&=&y^+,y;E)-\frac{\partial}{\partial x} G_0(x=y^-,y;E)=-\frac{2m}{\hbar^2},
\label{eq: matching}
\en
where the last equation is obtained from integrating both sides of Eq.~(\ref{eq: G0Eq}) around $x=y$.
The standard procedure of solving $G_0$ is assuming that it takes the following form~\cite{Jackson}:
\be
G_0(x,y;E)=A f_<\left(\min(x,y)\right) f_>\left(\max(x,y)\right),
\en
where $f_<(x)$ and $f_>(x)$ satisfy
\be
\left(\frac{\partial^2}{\partial x^2}-\alpha^2 x^2+2\alpha^2\varepsilon \right)f_{<,>}(x)=0,
\label{eq: DE}
\en
with the boundary conditions
\be
\lim_{x\to-\infty} f_<(x)=0,
\quad
\lim_{x\to\infty} f_>(x)=0,
\en
$f_>(x)$ can be expressed as combinations of two linearly independent solutions~\cite{Lebedev}:
\be
f_>(x)=c_1 e^{-\alpha^2 x^2/2} H_\nu(\alpha x)+c_2 e^{\alpha^2 x^2/2} H_{-\nu-1}(i\alpha x),
\label{eq:larger}
\en
with
\be
\nu\equiv\varepsilon-\frac{1}{2}.
\label{eq:muepsilon}
\en
From Eq.~(\ref{eq: DE}), we see that if $f(x)$ is a solution, $f(-x)$ is also a solution. It is convenient to use
\be
f_<(x)=c_3 e^{-\alpha^2 x^2/2} H_\nu(-\alpha x)+c_4 e^{\alpha^2 x^2/2} H_{-\nu-1}(-i\alpha x).
\label{eq:smaller}
\en
We see that both the real and imaginary parts of $e^{\alpha^2 x^2/2} H_{-\nu-1}(i\alpha x)$ 
are divergent as $x$ goes to $\pm \infty$. Nevertheless, we have
\be
\lim_{x\to -\infty} e^{-\alpha^2 x^2/2} H_\nu(-\alpha x)=0,
\quad
\lim_{x\to \infty} e^{-\alpha^2 x^2/2} H_\nu(\alpha x)=0.
\en
These boundary conditions lead to
\be
G_0(x,y;E)
&=&A e^{-\frac{\alpha^2(x^2+y^2)}{2}} 
H_{\varepsilon-1/2}\left(-\alpha \min(x,y)\right) H_{\varepsilon-1/2}\left(\alpha \max(x,y)\right),
\en
where we take $f_<(x)=\exp(-\alpha^2 x^2/2) 
H_\nu(-\alpha x)$ and $f_>(x)= \exp(-\alpha^2 x^2/2) H_\nu(\alpha x)$ in the above equation.
Substituting the above $G_0$ to the matching condition, Eq. (\ref{eq: matching}), we obtain
\be
G_0(x,y;E)&=&-\frac{2m}{\hbar^2}\frac{e^{-\frac{\alpha^2(x^2+y^2)}{2}} 
H_{\varepsilon-1/2}\left(-\alpha \min(x,y)\right) H_{\varepsilon-1/2}\left(\alpha \max(x,y)\right)}
{W\left(e^{-\alpha^2y^2/2} H_{\varepsilon-1/2}(-\alpha y),e^{-\alpha^2y^2/2}  H_{\varepsilon-1/2}(\alpha y)\right)},
\label{eq: G0W}
\en
where $W$ is the Wronskian assuring $G_0(x,y;E)$ to satisfy the matching condition (or discontinuity condition). 
Note that the Wronskian  in the above equation is a constant~\cite{Lebedev},
\be
W\left(e^{-\alpha^2y^2/2} H_{\varepsilon-1/2}(-\alpha y),e^{-\alpha^2y^2/2}  H_{\varepsilon-1/2}(\alpha y)\right)
&=&-\frac{2^{\varepsilon+1/2}\sqrt\pi\alpha}{\Gamma(\frac{1}{2}-\varepsilon)}.
\label{eq: WHnu}
\en
Finally we obtain the time-independent Green's function of a quantum SHO as
\be
G_0(x,y;E)=\frac{m  \Gamma(\frac{1}{2}-\varepsilon) }{2^{\varepsilon-1/2}\sqrt\pi \alpha\hbar^2} e^{-\frac{\alpha^2(x^2+y^2)}{2}} 
H_{\varepsilon-1/2}\left(-\alpha \min(x,y)\right) H_{\varepsilon-1/2}\left(\alpha \max(x,y)\right). 
\label{eq: G0}
\en
As we shall see that it is natural and straightforward to obtain eigenvalues and eigenfunctions from the above Green's function. 
Before we proceed further, we note that the Wronskian of $f_<$ and $f_>$ is proportional to the Wronskian of $H_\nu(-\alpha x)$ and $H_\nu(\alpha x)$, and
$H_\nu(-\alpha x)$ and $H_\nu(\alpha x)$ are linearly independent for $\nu\neq 0,1,\cdots$~\cite{Lebedev}.
On the contrary for $\nu=n=0,1,2,3\cdots$, since the Hermite polynomials 
have the following property:
\be
H_n(-\alpha x)=(-1)^n H_n(\alpha x),
\label{eq: Hn}
\en
$H_{\nu=n}(-\alpha x)$ and $H_{\nu=n}(\alpha x)$ are linearly dependent, and hence the corresponding Wronskian is vanishing.
We will return to this later.

Note that it is easy to see that $G_0(x,y;E)$ satisfies the following relations:
\be
G_0(x,y;E)=G_0(y,x;E)=G_0(-x,-y; E)=G_0(-y,-x,E),
\label{eq: G0 relations}
\en
which echo the very same properties of the more familiar time-dependent Green's function $K(x,t; y,t_0)$ (see for example \cite{Sakurai})
\be
K(x,t;y,t_0)=\left(\frac{m\omega}{2\pi i\hbar \sin\omega (t-t_0)}\right)^{1/2}
\exp\left(\frac{im\omega}{2\hbar\sin\omega (t-t_0)}[(x^2+y^2)\cos\omega(t-t_0)-2 xy]\right),
\en
which is related to $G_0(x,y,E)$ through a Fourier transform.
\begin{figure}[t!]
\centering
\captionsetup{justification=raggedright}
\subfigure[]{
  \includegraphics[width=6cm]{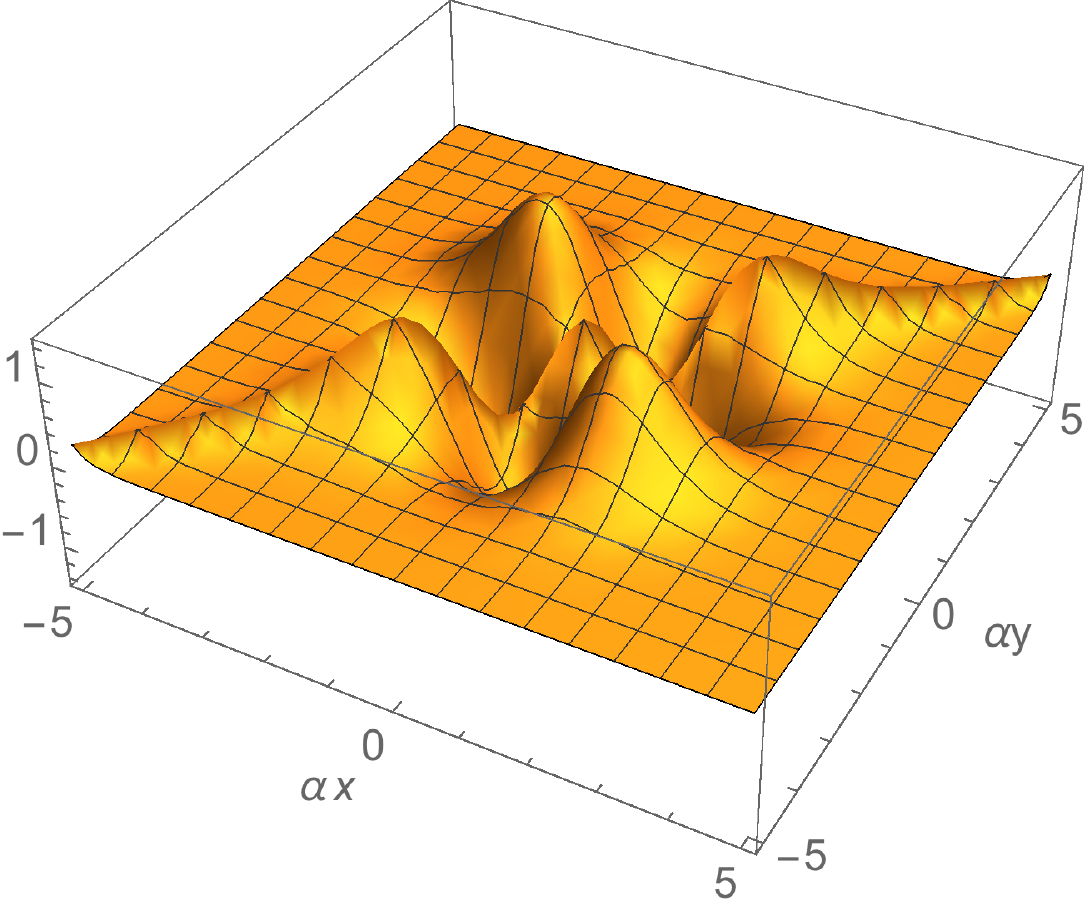}
 }
 \hspace{0.5cm}
\subfigure[]{
  \includegraphics[width=6cm]{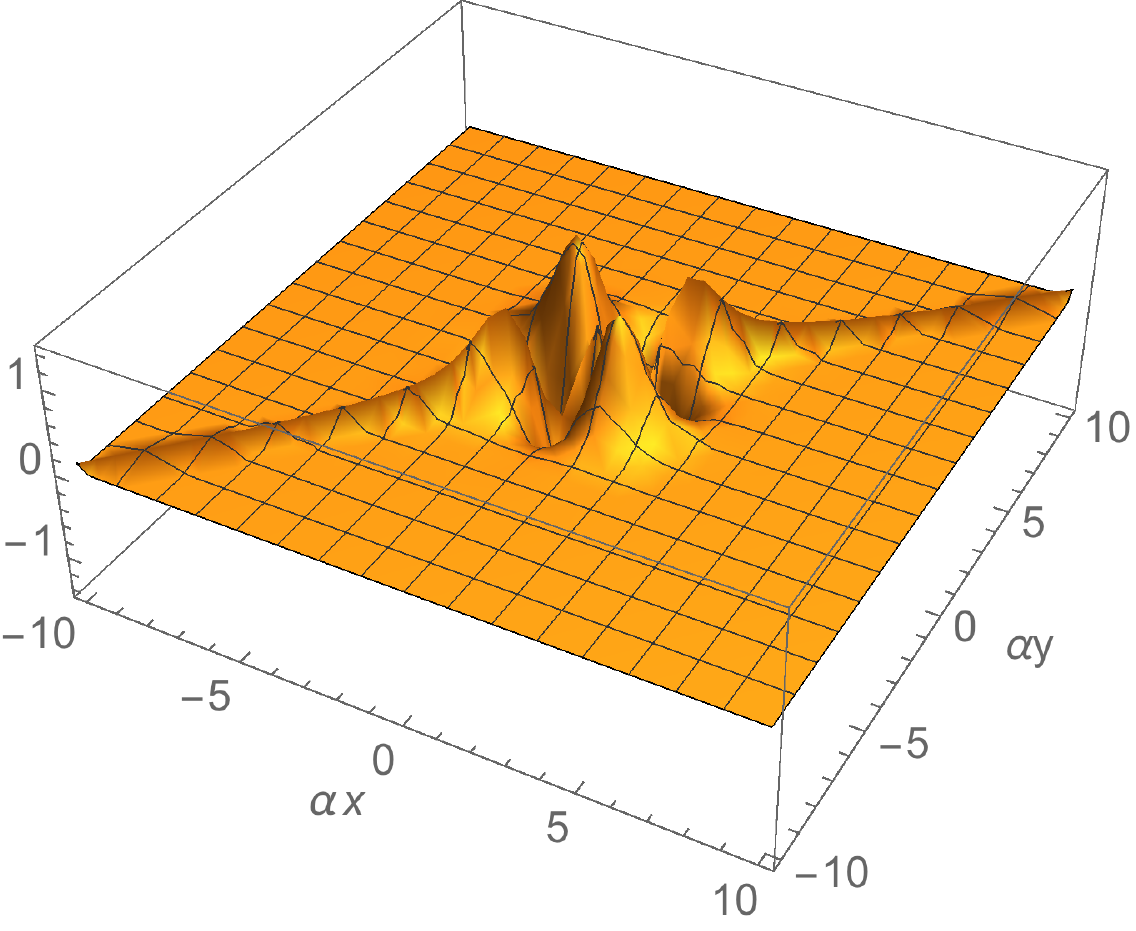}
}
\subfigure[]{
  \includegraphics[width=6cm]{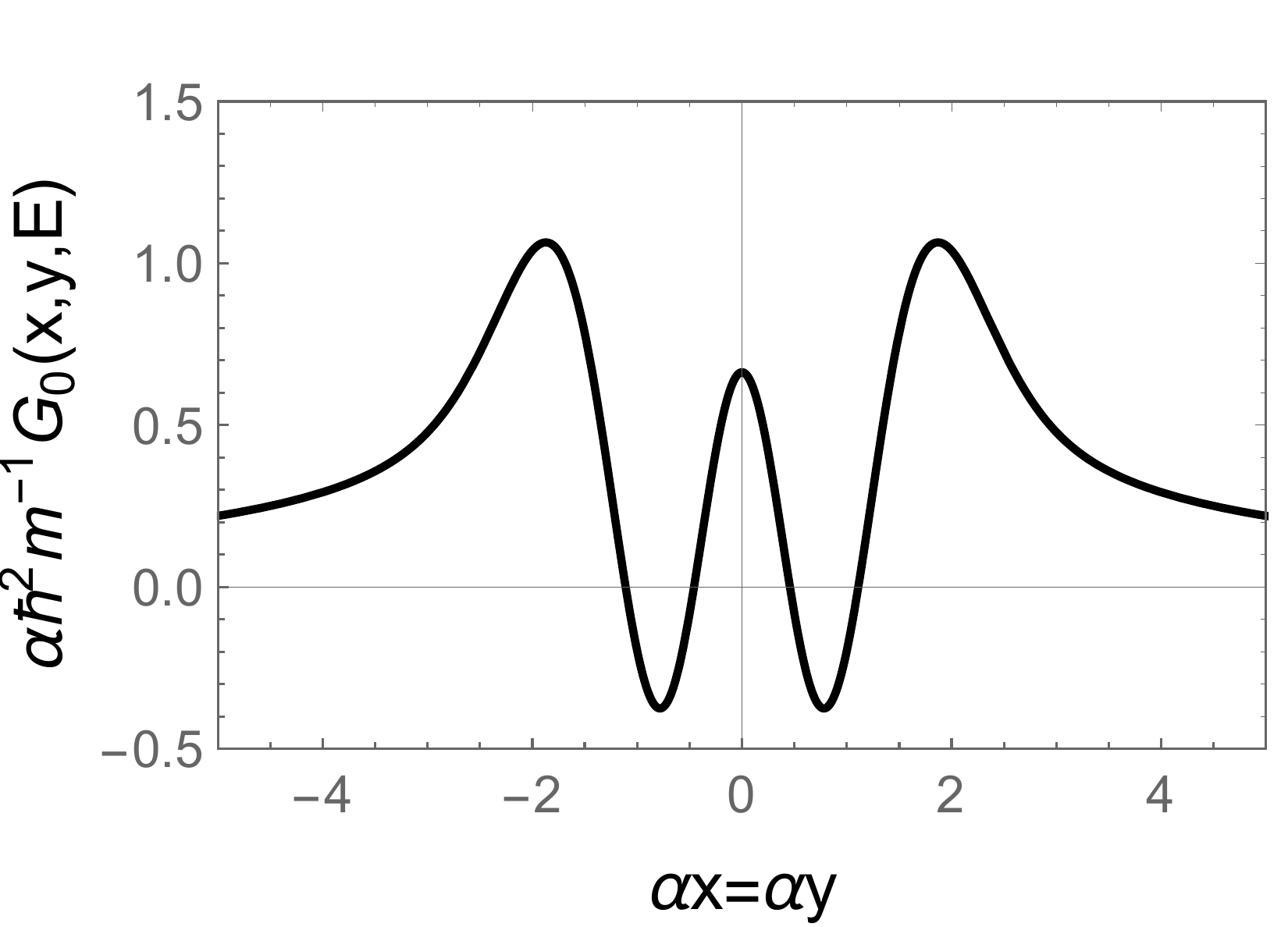}
}
\hspace{0.5cm}
\subfigure[]{
  \includegraphics[width=6cm]{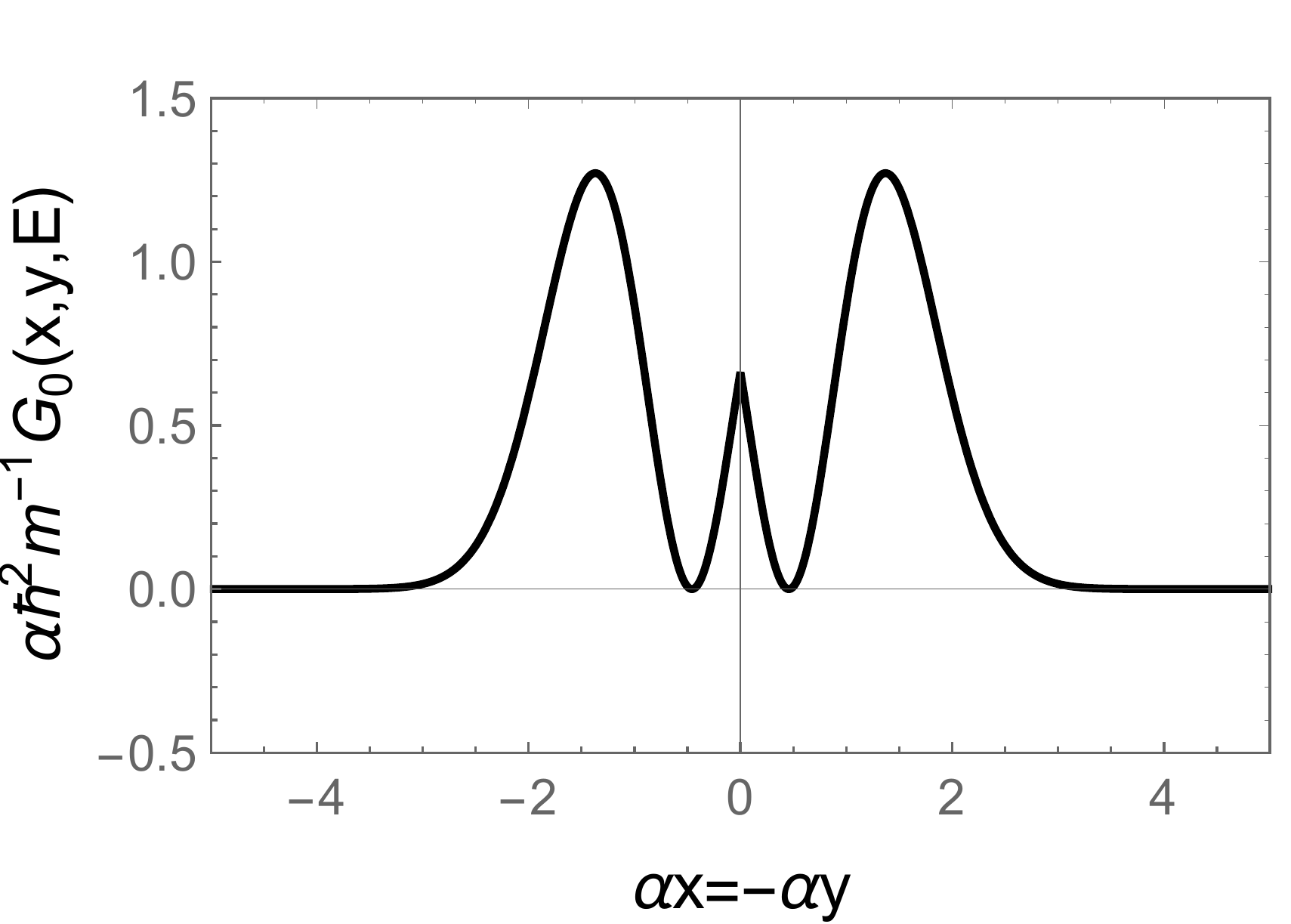}
}
\caption{3D plot of 
 $\alpha\hbar^2 m^{-1} G_0(x,y, E)$ using Eq. (\ref{eq: G0}) with $\varepsilon=2.1$
 for (a) $-5<\alpha x,\alpha y<5$ and (b) $-10<\alpha x,\alpha y<10$, and the section views  
 along the line $x=y$ in (c) and the line $x=-y$ in (d), respectively.} 
 \label{fig: G0}
\end{figure}

For illustration, we plot $G_0(x,y,E)$ for $\varepsilon=2.1$  in Fig.~\ref{fig: G0}. 
Note that the relations stated in Eq.~(\ref{eq: G0 relations}) can be seen from the plots, where the graphs are clearly both mirror symmetric to the $x=y$ and the $x=-y$ lines. 
It is interesting to see that $G_0$ 
has some activities around the origin and has two long (but damping) tails along the $x=y$ line (in the positive $x,y$ and the negative $x,y$ directions) and remains highly suppressed in other region. 

The energy eigenvalues and eigenvectors can be easily obtained from the above $G_0$.
It is well known that the Gamma function $\Gamma(\frac{1}{2}-\varepsilon)$ has poles at 
\be
\frac{1}{2}-\varepsilon=-n\equiv\frac{1}{2}-\varepsilon_n,
\quad
n=0,1,2,3,\dots,
\label{eq: E0n}
\en
giving the well known result:
\be
E_n=\frac{\alpha^2\hbar^2}{m}\varepsilon_n=(n+\frac{1}{2})\hbar\omega,
\en 
and from  
\be
\lim_{\varepsilon\to\varepsilon_n}(\varepsilon_n-\varepsilon) \Gamma(\frac{1}{2}-\varepsilon)=\frac{(-1)^n}{\Gamma(n+1)},
\en
we obtain
\be
\lim_{E\to E_n} (E_n-E)G_0(x,y,;E)
&=&\frac{\alpha}{2^{n}\sqrt\pi\,\Gamma(n+1)}
e^{-\frac{\alpha^2(x^2+y^2)}{2}} 
H_n\left(\alpha x\right) H_n\left(\alpha y\right),
\en
which is just $u_n(x) u^*_n(y)$, where $u_n(x)$ is the usual time-independent wave function of simple harmonic oscillator
\be
u_n(x)=\eta\left(\frac{\alpha}{2^{n}\sqrt\pi\,\Gamma(n+1)}\right)^{1/2}
e^{-\frac{\alpha^2 x^2}{2}} 
H_n\left(\alpha x\right),
\en
with $\eta$ is a phase factor conventionally taken to be 1.

In summary the poles of $G_0$ are determined from the zeros of the Wronskian, $W(f_<(x),f_>(x))$. 
As noted for $\nu\neq n$ the Wronskian is not vanishing, while for $\nu=n$, $f_<(x)$ and $f_>(x)$ are linearly dependent as shown in Eq.~(\ref{eq: Hn}). Therefore $\nu\equiv\varepsilon-1/2=n$ are the zeros of the Wronskian and hence the poles of the Green's function.

A related observation at the wave function level was given in ref.~\cite{Viana-Gomes}. We can redo their argument using the technic similar to the above derivation and give further insight into the problem.
We express the wave function $u_\nu(x)$ as
\be
u_\nu(x)&=& N  e^{-\frac{\alpha^2(x^2+y^2)}{2}} 
H_{\nu}\left(-\alpha \min(x,y)\right) H_{\nu}\left(\alpha \max(x,y)\right),
\label{eq: unu}
\en
with $\nu=\varepsilon-1/2$, $N$ a normalization factor, and an arbitrary $y$ as long as $H_{\nu}(\pm\alpha y)$ is non-vanishing. 
By construction the wave function is guaranteed to be continuous at $x=y$ and is the solution of the Schr\"odinger equation (for $x\neq y$)
\be
\left(-\frac{\hbar^2}{2m}\frac{\partial^2}{\partial x^2}+\frac{1}{2} m\omega^2 x^2 \right)u_\nu(x)=E u_\nu(x),
\en
satisfying the boundary conditions
\be
\lim_{x\to\pm\infty} u_\nu(x)=0,
\en
for any (real) value of $\varepsilon=\nu+1/2$.
As pointed out in ref.~\cite{Viana-Gomes}, in contrary to the usual treatment in most text books, the satisfactions of these requirements do not necessarily lead to a viable wave function solution.
A viable solution of the above Schr\"{o}dinger equation (including $x=y$) requires the derivative of the wave function to be continuous as well~\cite{Viana-Gomes}.
It can be easily seen that this requires, at the matching point ($x=y$), that we must have
\be
0=\frac{1}{ N}u'_\nu(x)\Big |^{y^+}_{y^-}=W(f_<(y), f_>(y))
=e^{-\alpha y^2} W\left(H_\nu(-\alpha y), H_\nu(\alpha y)\right)
=\frac{2^{\varepsilon-1/2}\sqrt\pi \alpha}{\Gamma(\frac{1}{2}-\varepsilon) },
\label{eq: u'}
\en
giving $\nu=\varepsilon-1/2=n$, 
and
\be
u_n(x)=N'   e^{-\frac{\alpha^2 x^2}{2}}
H_n\left(\alpha x\right),
\en
where $N'\equiv (-1)^n N\exp(-\alpha^2 y^2/2) H_n(\alpha y)$.
Note that this happens only at $\nu=n$, where $H_\nu(-\alpha x)$ and $H_{\nu}(\alpha x)$ are linearly dependent [see Eq. (\ref{eq: Hn})] and produce a vanishing Wronskian.

In fact, it is exactly the same place and the very same source where the Green's function $G_0$ has poles in $E$ [see Eqs. (\ref{eq: G0W}) and (\ref{eq: WHnu})]. 
At a first sight it seems that this is just a coincident, as the Wronskian in $G_0$ is to assure the Green's function to satisfy the matching condition or discontinuity condition, Eq.~(\ref{eq: matching}), while the Wronskian in $u_\nu$ is to govern the continuity of the wave function at $y$ and is required to be vanishing to give a viable wave function. 
To see the connection we rewrite $G_0(x,y;E)$ in Eq.~(\ref{eq: G0W}), with the help of Eq. (\ref{eq: unu}), as
\be
G_0(x,y;E)&=&-\frac{2m}{\hbar^2}\frac{u_\nu(x)/N}
{W\left(e^{-\alpha^2y^2/2} H_{\varepsilon-1/2}(-\alpha y),e^{-\alpha^2y^2/2}  H_{\varepsilon-1/2}(\alpha y)\right)}.
\label{eq: GWu}
\en
and,
consequently,
\be
\Delta G'_0\equiv\frac{\partial}{\partial x}G_0(x,y;E)\Big |^{x=y^+}_{x=y^-}&=&-\frac{2m}{\hbar^2 N}\frac{u'_\nu(x)\big |^{y^+}_{y^-}}
{W\left(e^{-\alpha^2y^2/2} H_{\varepsilon-1/2}(-\alpha y),e^{-\alpha^2y^2/2}  H_{\varepsilon-1/2}(\alpha y)\right)}.
\label{eq: GWuD}
\en
The matching condition or discontinuity condition requires the derivative of $G_0$ to be discontinuous at $x=y$, i.e. $\Delta G'_0=-2m/\hbar\neq0$.
However, as $\varepsilon$ approaching $n+1/2$, the derivative of the numerator of $G_0$ in Eq.~(\ref{eq: GWu}) tends to be continuous at $x=y$ [see Eq. (\ref{eq: u'})],
leading to a vanishing numerator in $\Delta G'_0$ [see Eq.~(\ref{eq: GWuD})].
The Wronskian in the denominator of $\Delta G'_0$ keeps doing it's job to assure $G'_0$ to satisfy the discontinuity condition by balancing the Wronskian from the derivative of the numerator.
Hence a continuous $u'_\nu$ at $x=y$ corresponds to a vanishing Wronskian in the denominator in $G_0$.
The Wronskians in the numerator and the denominator of $\frac{\partial}{\partial x}G_0(x,y;E)\big |^{x=y^+}_{x=y^-}$ as shown in the above equation are the very same Wronskian as required from the matching condition Eq.~(\ref{eq: matching}). 
In short, the poles of $G_0$ in $E$ is tied to the continuity of $u'_\nu$ at $x=y$.

We see that the Wronskian plays some interesting and important roles in $G_0$ and $u_\nu$. 
Although it is possible to obtain $G_0(x,y; E)$ by performing the Fourier transform of 
$K_0(x,t;y,t_0)$ using an integral of Bessel function~\cite{Kleinert:2004ev}, the interesting point we see in the above discussion will be obscured. 

\section{Time-independent Green's function of a quantum simple harmonic oscillator with a generic delta-function potential}

\begin{figure}[t!]
\centering
\captionsetup{justification=raggedright}
\subfigure[]{
  \includegraphics[width=7cm]{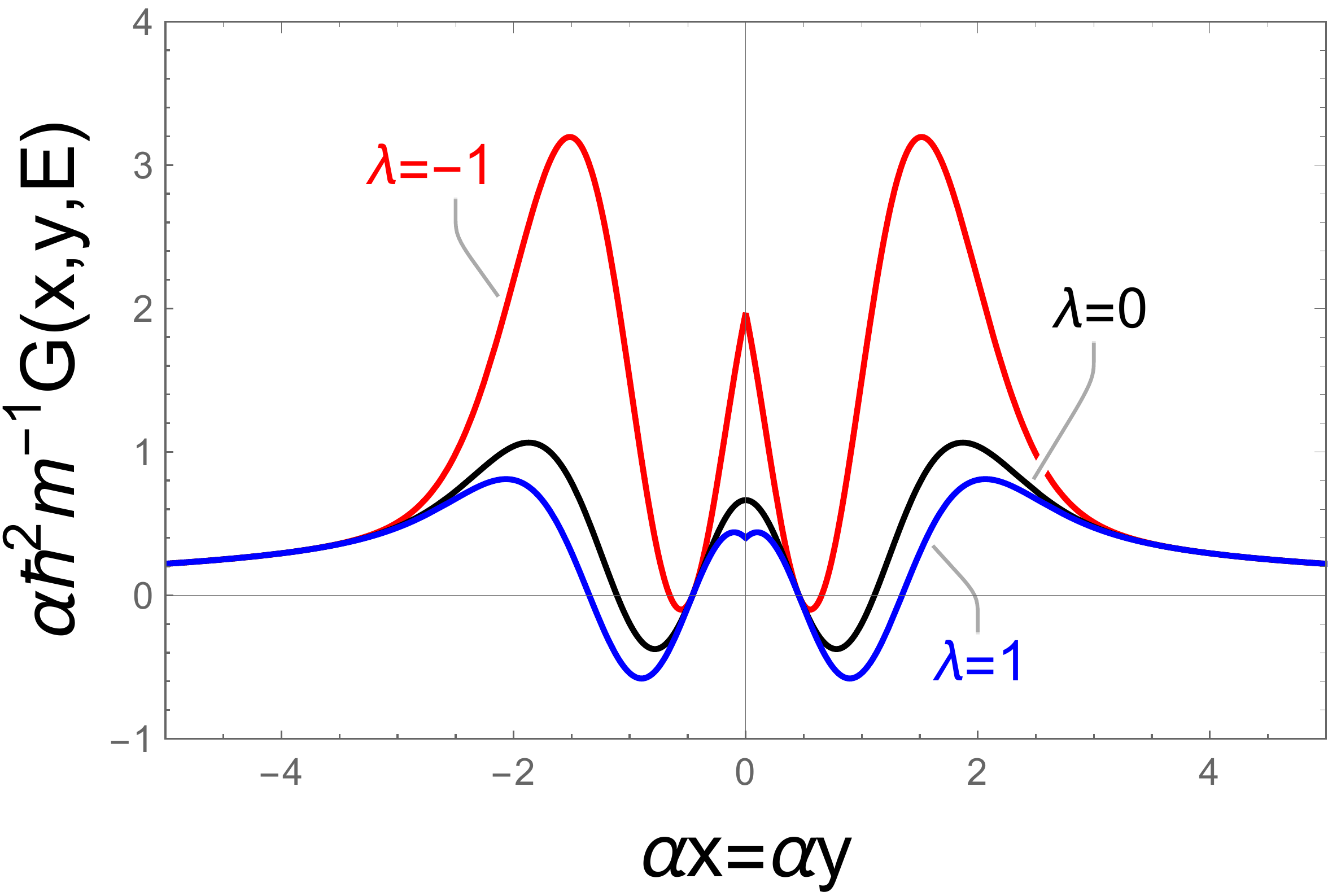}
}
\hspace{0.5cm}
\subfigure[]{
  \includegraphics[width=7cm]{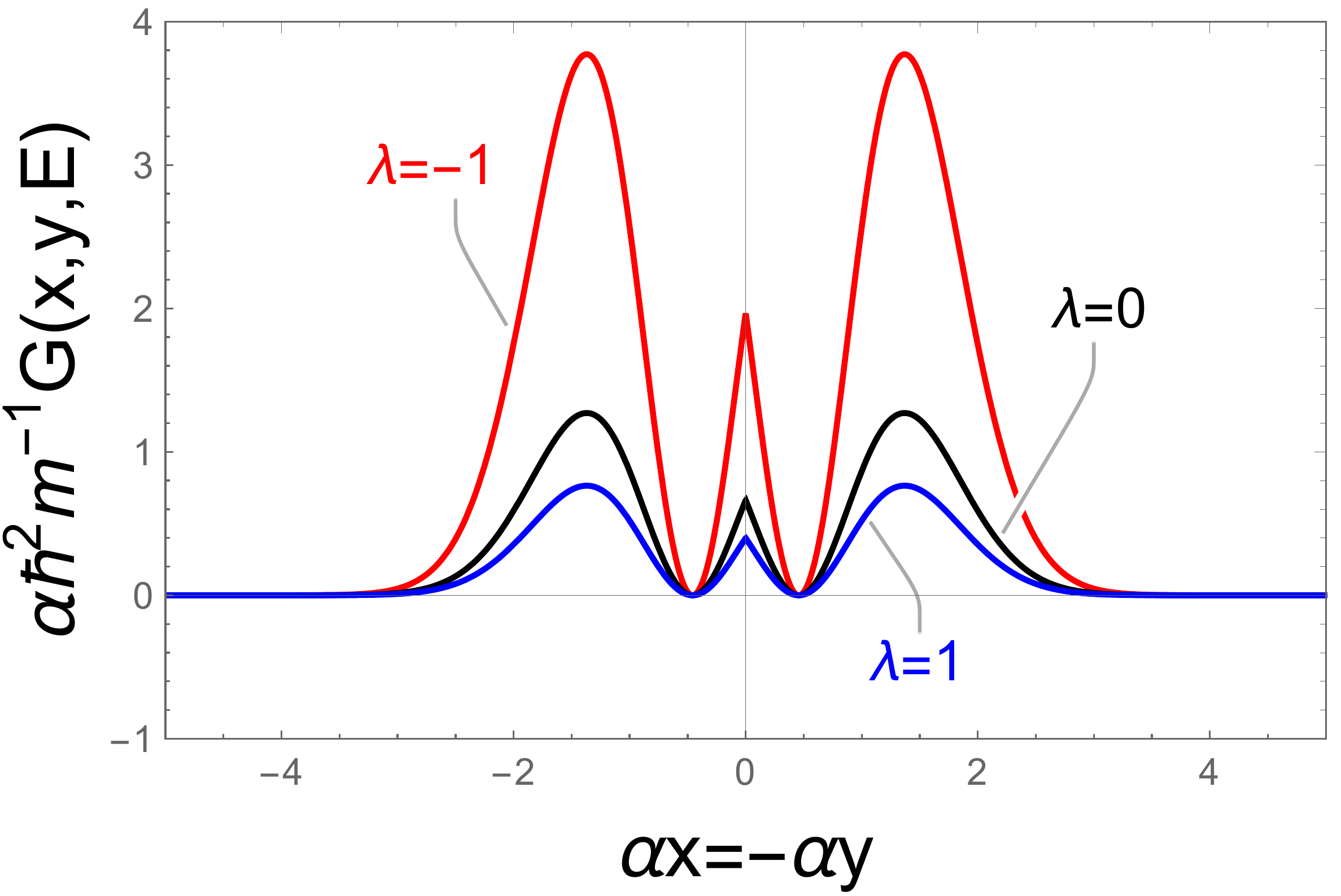}
}
\caption{
  Section views of $\alpha\hbar^2 m^{-1} G(x,y; E)$ with $V_1(x)=\frac{\alpha\hbar^2}{m}\lambda\delta(x) $ and $\varepsilon=2.1$ along the line $x=y$ in (a) and the line $x=-y$ in (b).} 
 \label{fig: Glam3}
\end{figure}

In this section, we follow the approach of ref.~\cite{Cavalcanti} 
to obtain the Green's functions for SHO with an additional delta-function potential.
Let $H_0$, $G_0$ and $H$, $G$ be the Hamiltonian and the Green's function of the quantum systems of pure SHO and SHO with a 
delta-function potential $V_1(x)=\frac{\alpha\hbar^2}{m}\lambda\delta(x-a)$, respectively. Hence we have
\be
 (H_0-E)G_0(x,y;E)= \delta(x-y),
 \quad
 (H-E)G(x,y;E)= \delta(x-y),
\en
where 
$H=H_0+V_1$.
In the above, the second equation can be rewritten as 
\be
(H_0-E)G(x,y;E)= \delta(x-y)-V_1(x)G(x,y;E).
\en
Multiplying $G_0(y,x)$ on both sides and integrating them with respective to $x$ from $-\infty$ to $\infty$, we obtain the following relation:
\be 
G(x,y;E) &=&G_0(x,y;E)-\frac{\alpha\hbar^2}{m}G_0(x,a;E)G(a,y;E),
\label{eq:GG0}
\en
which gives~\cite{Cavalcanti}
\be
G(x,y;E)=G_0(x,y;E)-\frac{G_0(x,a;E)G_0(a,y;E)}{\frac{1}{\frac{\alpha\hbar^2}{m}\lambda}+G_0(a,a;E)}.
\en
Using Eq.~(\ref{eq: G0}), the Green's function of the quantum system of SHO with a delta-function potential 
$V_1(x)=\frac{\alpha\hbar^2}{m}\lambda\delta(x-a)$ can be written explicitly as
\be
G(x,y;E)
&=&\frac{m\Gamma (\frac{1}{2}-\varepsilon )}{{{2}^{\varepsilon -1/2}}\sqrt{\pi }\alpha {{\hbar }^{2}}}{{e}^{-\frac{{{\alpha }^{2}}({{x}^{2}}+{{y}^{2}})}{2}}}{{H}_{\varepsilon -1/2}}\left( -\alpha \min (x,y) \right){{H}_{\varepsilon -1/2}}\left( \alpha \max (x,y) \right)
\non\\
&&
-\frac{{{\left( \frac{m\Gamma (\frac{1}{2}-\varepsilon )}{{{2}^{\varepsilon -1/2}}\sqrt{\pi }\alpha {{\hbar }^{2}}} \right)}^{2}}{{e}^{-\frac{{{\alpha }^{2}}({{x}^{2}}+{{y}^{2}})}{2}}}}{\frac{m}{\lambda \alpha {{\hbar }^{2}}}+\frac{m\Gamma (\frac{1}{2}-\varepsilon )}{{{2}^{\varepsilon -1/2}}\sqrt{\pi }\alpha {{\hbar }^{2}}}H_{\varepsilon -1/2}( -\alpha a)H_{\varepsilon -1/2}(\alpha a)}
\non\\
&&\quad
\times 
{{H}_{\varepsilon -1/2}}\left( -\alpha \min (x,a) \right){{H}_{\varepsilon -1/2}}\left( \alpha \max (x,a) \right)
\non\\
&&\quad
\times
{{H}_{\varepsilon -1/2}}\left( -\alpha \min (a,y) \right){{H}_{\varepsilon -1/2}}\left( \alpha \max (a,y) \right).
\label{eq:GxyE}
\en
For illustration we plot the section views of $G(x,y;E)$ in the case of $a=0$ with $\varepsilon=2.1$ along the line $x=y$ in Fig.~\ref{fig: Glam3}(a) and the line $x=-y$ in Fig.~\ref{fig: Glam3}(b), respectively. 
We can see clearly the discontinuity of first derivative of $G(x,y;E)$ at origin for $\lambda=\mp 1$ in the plots.

It is well known that by analysing the pole of $G(x,y;E)$ in $E$ one can obtain the energy levels $E_n$.
From Eq.~(\ref{eq:GxyE}), 
by taking $x=y=a\neq 0$, we have
\be 
G(a,a;E)=\frac{m e^{-\alpha^2a^2}H_{\varepsilon-1/2}(-\alpha a)H_{\varepsilon-1/2}(\alpha a) }
{\alpha\hbar^2\bigg( \frac{\sqrt \pi 2^{\varepsilon -1/2}}{\Gamma(1/2-\varepsilon)}
+\lambda e^{-\alpha^2a^2}H_{\varepsilon-1/2}(-\alpha a)H_{\varepsilon-1/2}(\alpha a) \bigg)},
\label{eq: Gaa}
\en
we can obtain the following 
transcendental equation from the poles of the above function:
\be
\frac{2^{\varepsilon+1/2}\sqrt\pi}{\Gamma(\frac{1}{2}-\varepsilon)}+
2\lambda e^{-\alpha^2a^2}H_{\varepsilon-1/2}(-\alpha a)H_{\varepsilon-1/2}(\alpha a)=0.
\label{eq: dis1}
\en
It can be easily seen that in the $\lambda=0$ limit, we return to the familiar energy level of the SHO case [see Eq. (\ref{eq: E0n})].

\begin{figure}[t!]
\centering
\captionsetup{justification=raggedright}
  \includegraphics[width=0.6\textwidth]{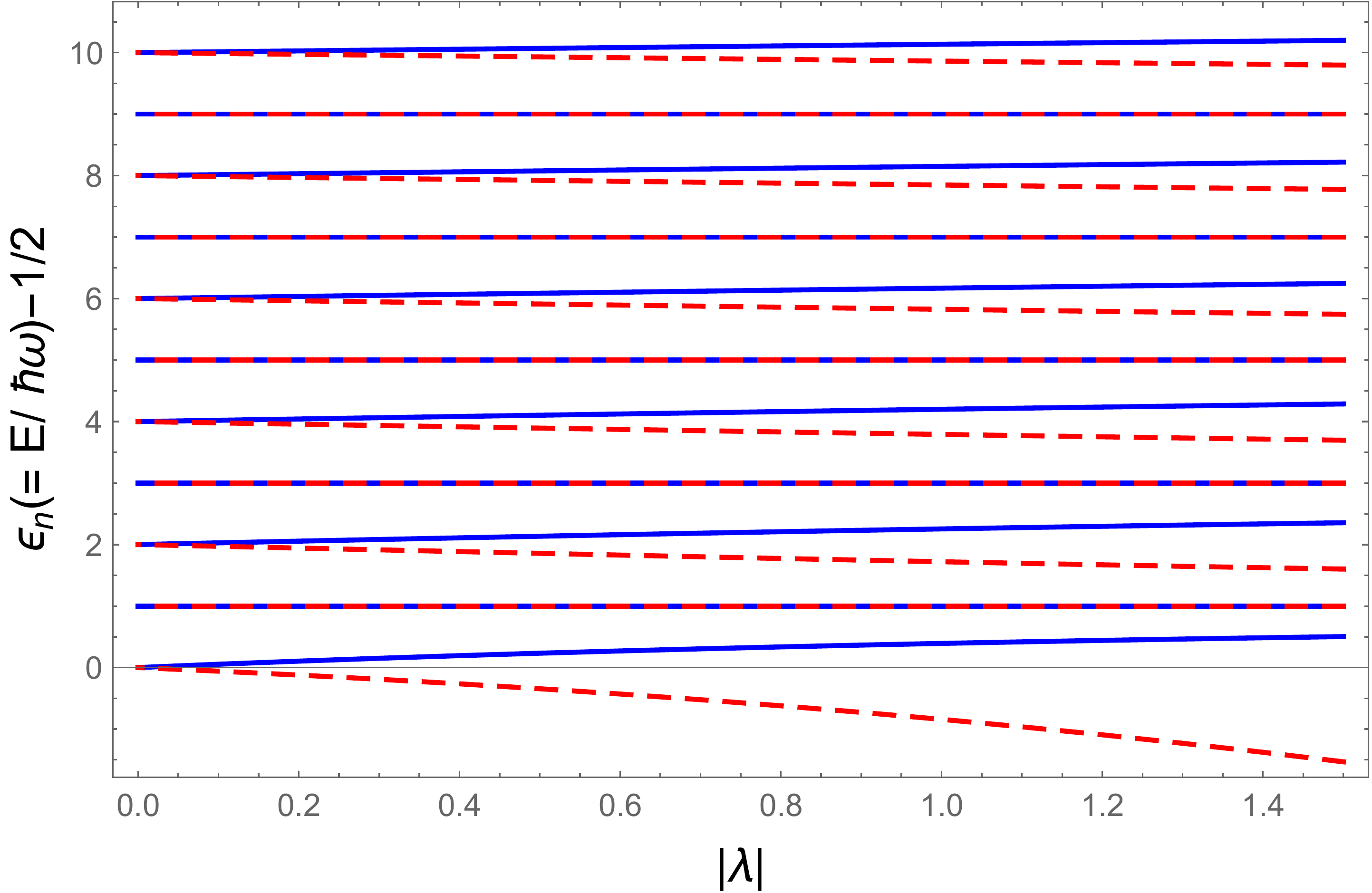}
  \caption{The plot of energy eigenvalue $\varepsilon_n-1/2$ versus the coupling strength $\lambda$ is shown. The solid and the dashed line correspond with $\lambda>0$ and $\lambda<0$, respectively.}
  \label{fig:GEnlambda}
\end{figure}

We can compare our result with the one in ref.~\cite{Viana-Gomes} by taking the $a\to 0$ limit.~\footnote{In fact the analysis of Eq. (\ref{eq: Gaa}) is not applicable for the odd level case, since the numerator will go to zero as well. A more complete analysis is needed and interestingly the result in Eq. (\ref{eq: dis1}) still holds.} 
Fig.~\ref{fig:GEnlambda} shows the plot of the first eleven energy eigenvalues $\varepsilon_n-1/2$ $(n=0,1,\cdots,10)$ versus the coupling strength $\lambda$ in the region $-1.5\leq\lambda\leq 1.5$. For $\lambda=0$ or $n$ is odd, the energy eigenvalues go back to those in the pure quantum SHO system.
This can be easily understood: since in the presence of the $\delta(x)$ potential, the full potential in the Schr\"odinger equation is still parity even giving parity even and odd solutions, the energy levels of the parity odd states are unaffected by the $\delta(x)$ potential 
as the odd wave function is vanishing at the origin.
For $n$ is even, we see that the energy eigenvalue increases with increasing $\lambda$, but it never crosses over the adjacent eigenvalues of odd energy levels. 
To be specific, we show the numerical values of the first six even parity eigenvalues $\varepsilon_n-1/2$ for $\lambda=0, \mp0.5$ and $\mp 1$
in Table~\ref{tab:GdeltaEn}. Our results agree well with those in ref.~\cite{Viana-Gomes}.

\begin{table}[t!]
\centering
{\begin{tabular}{|c|c|c|c|c|c|c|}
\hline
$\varepsilon_n-1/2$
          & $n=0$
          & $n=2$
          & $n=4$          
          & $n=6$
          & $n=8$
          & $n=10$
           \\
           \hline
$\lambda=0$
          &0
          &2
          &4 
          &6
          &8
          &10 
           \\     
           \hline
$\lambda=-0.5$
          &-0.344434
          & 1.85734
          & 3.89395
          & 5.91181
          & 7.92289
          & 9.93062
           \\     
$\lambda=0.5$          
          & 0.233518
          & 2.13541
          & 4.10367
          & 6.08703
          & 8.07642
          & 10.0689
           \\     
           \hline 
$\lambda=-1$  
          &-0.842419
          & 1.72077
          & 3.79123
          & 5.82578
          & 7.84733
          & 9.86242
           \\     
$\lambda=1$  
          & 0.392744
          & 2.25464
          & 4.2002 
          & 6.16991
          & 8.15009
          & 10.1359
           \\     
           \hline                                                                                                                                                  
\end{tabular}
}
\caption{The first six even parity eigenvalues $\varepsilon_n-1/2$ for $\lambda=0,\mp 0.5,\mp 1$.}
\label{tab:GdeltaEn}
\end{table}

\section{Solutions of Schr\"odinger equation of a simple harmonic oscillator system with a generic 
delta-function potential}

In this section, the similar technics previously used to solve the Green's function can also be applied to solving the wave function directly from the Schr\"odinger equation. We first write down the Schr\"odinger equation for the quantum system of SHO with a generic delta-function potential: 
\be
\left(-\frac{\hbar^2}{2m}\frac{\partial^2}{\partial x^2}+V(x) \right)v (x; a)=E v(x;a),
\en
where
\be
V(x)=\frac{1}{2}m\omega^2 x^2+\frac{\alpha\hbar^2}{m}\lambda\delta(x-a)=V_0(x)+V_1(x).
\en
The wave function needs to satisfy the boundary conditions:
\be
\lim_{x\to\pm\infty}v(x;a)=0,
\en
and the matching conditions around $x=a$:
\be
v(x=a^+;a)=v(x=a^-;a),
\en
\be
\frac{\partial}{\partial x} v(x=a^+;a)-\frac{\partial}{\partial x} v(x=a^-;a)=2\alpha\lambda v(a;a).
\label{eq: matchingv}
\en
Using the same technics in Sec. II, we assume that  
the wave function is of the form:
\be
v_\nu(x;a)
&=& N  e^{-\frac{\alpha^2(x^2+a^2)}{2}} 
H_{\nu}\left(-\alpha \min(x,a)\right) H_{\nu}\left(\alpha \max(x,a)\right),
\label{eq: vy1form}
\en
where $N$ is the the normalization constant and $\nu=\varepsilon-1/2$.

\begin{table}[b!]
\centering
\captionsetup{justification=raggedright}
{\begin{tabular}{|c|c|c|c|c|c|c|}
\hline
$\varepsilon_n-1/2$
          & $n=0$
          & $n=1$
          & $n=2$          
          & $n=3$
          & $n=4$
          & $n=5$
           \\
           \hline
$\lambda=0$
          & 0
          & 1
          & 2          
          & 3
          & 4
          & 5
           \\
           \hline
$\lambda=-0.5$
          &-0.288982
          & 0.895074
          & 1.97192
          & 2.88647
          & 3.99943
          & 4.90350
           \\     
           \hline 
$\lambda=0.5$          
          & 0.16908
          & 1.10823
          & 2.02645
          & 3.11238
          & 4.00057
          & 5.09438
           \\     
           \hline 
$\lambda=-1$  
          &-0.750901
          & 0.809830
          & 1.954355
          & 2.78069
          & 3.99885
          & 4.80935
           \\     
           \hline
$\lambda=1$  
          & 0.267782
          & 1.20385
          & 2.05035
          & 3.21619
          & 4.00114
          & 5.18277
           \\     
           \hline                                                                                                                                                  
\end{tabular}
}
\caption{The first six exact and perturbed energy eigenvalues $\varepsilon_n-1/2$ for $\alpha a=0.5$ with 
$\lambda=0,\mp 0.5, \mp 1$ are shown.}
\label{tab:WdeltaEn}
\end{table}
\begin{figure}[t!]
\centering
\captionsetup{justification=raggedright}
\subfigure[\ $n=0$ 
]{
  \includegraphics[width=6.5cm]{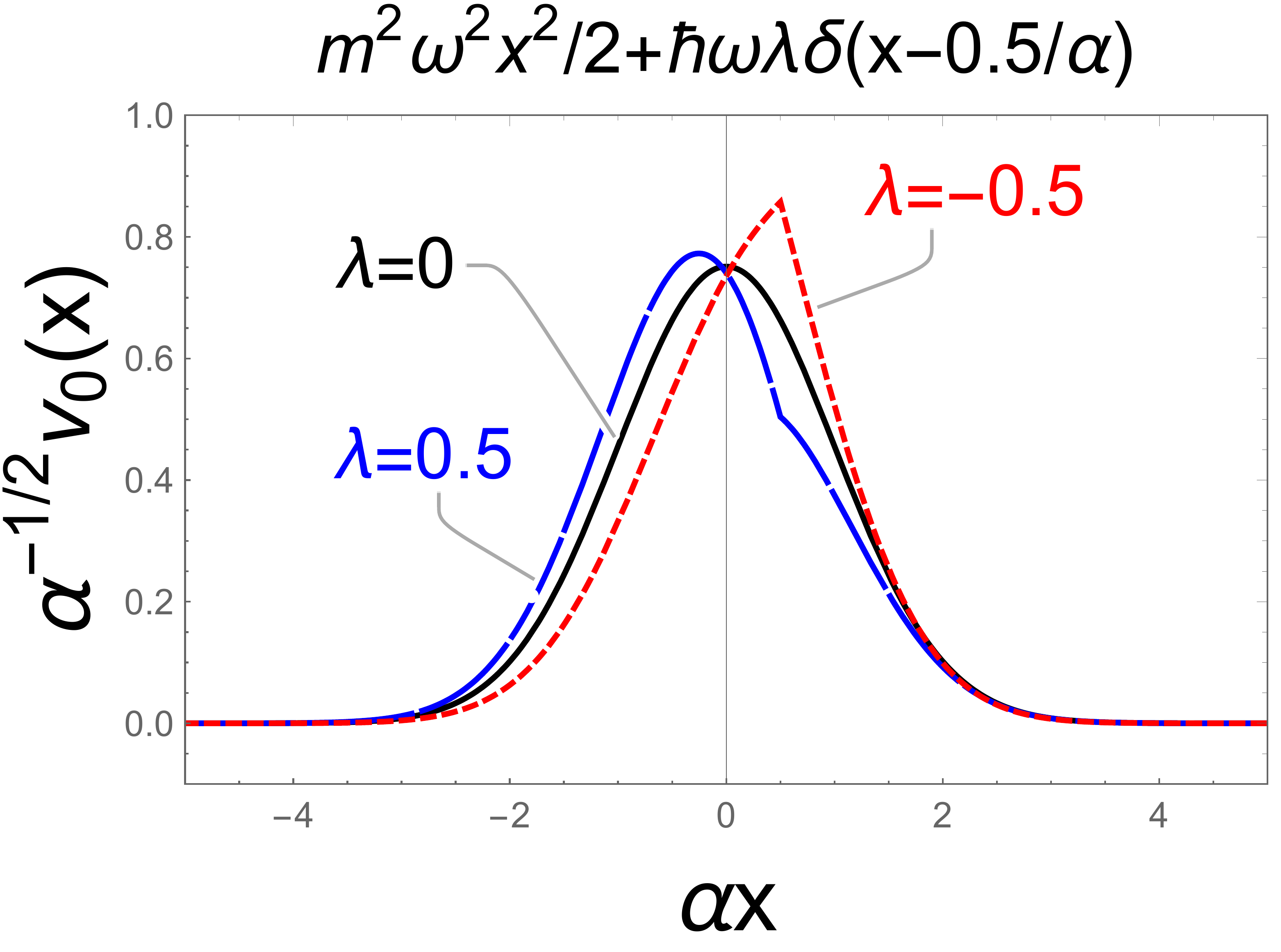}
}
\subfigure[\ $n=1$ 
]{
  \includegraphics[width=6.5cm]{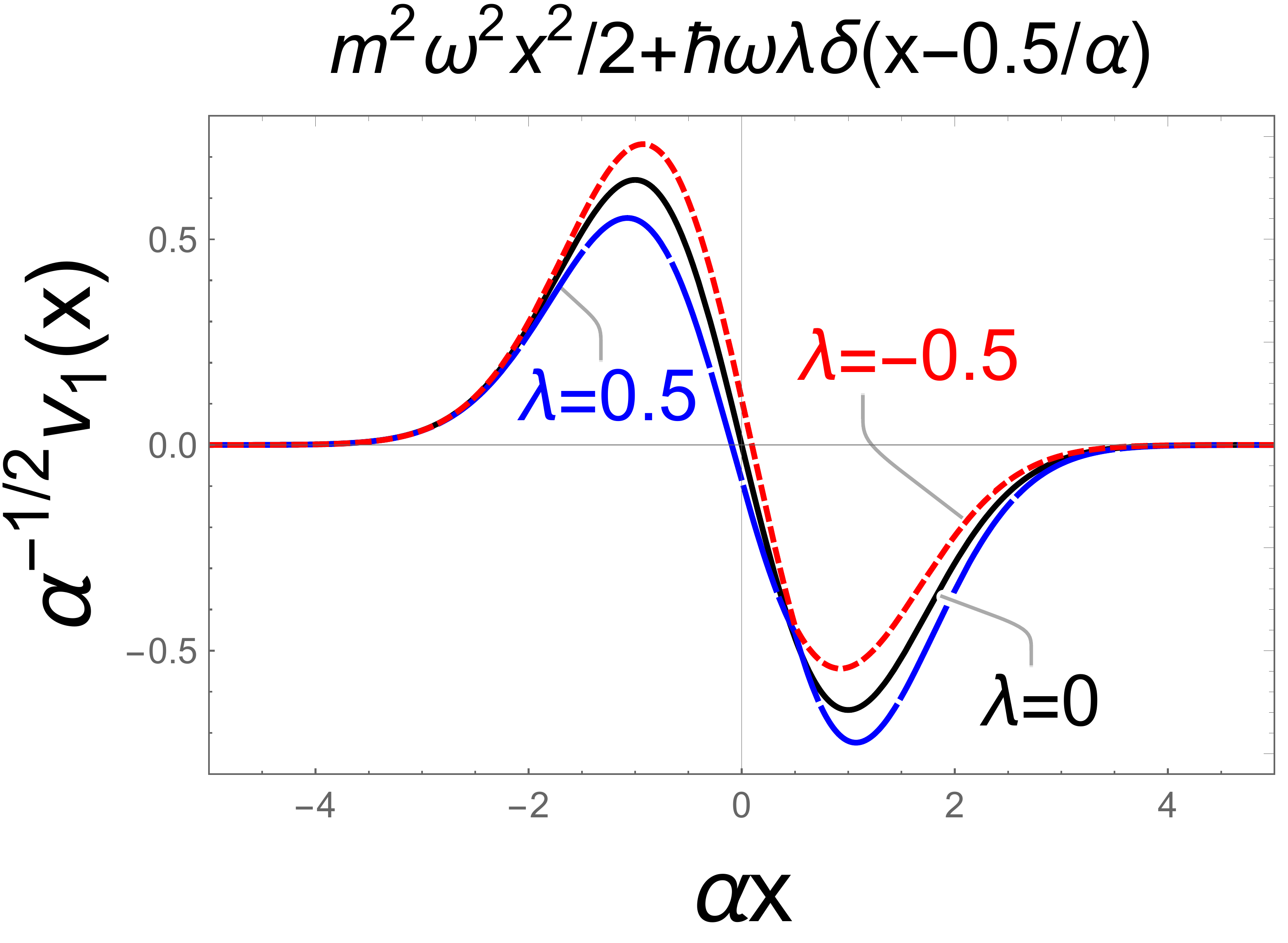}
}
\subfigure[\ $n=2$
]{
  \includegraphics[width=6.5cm]{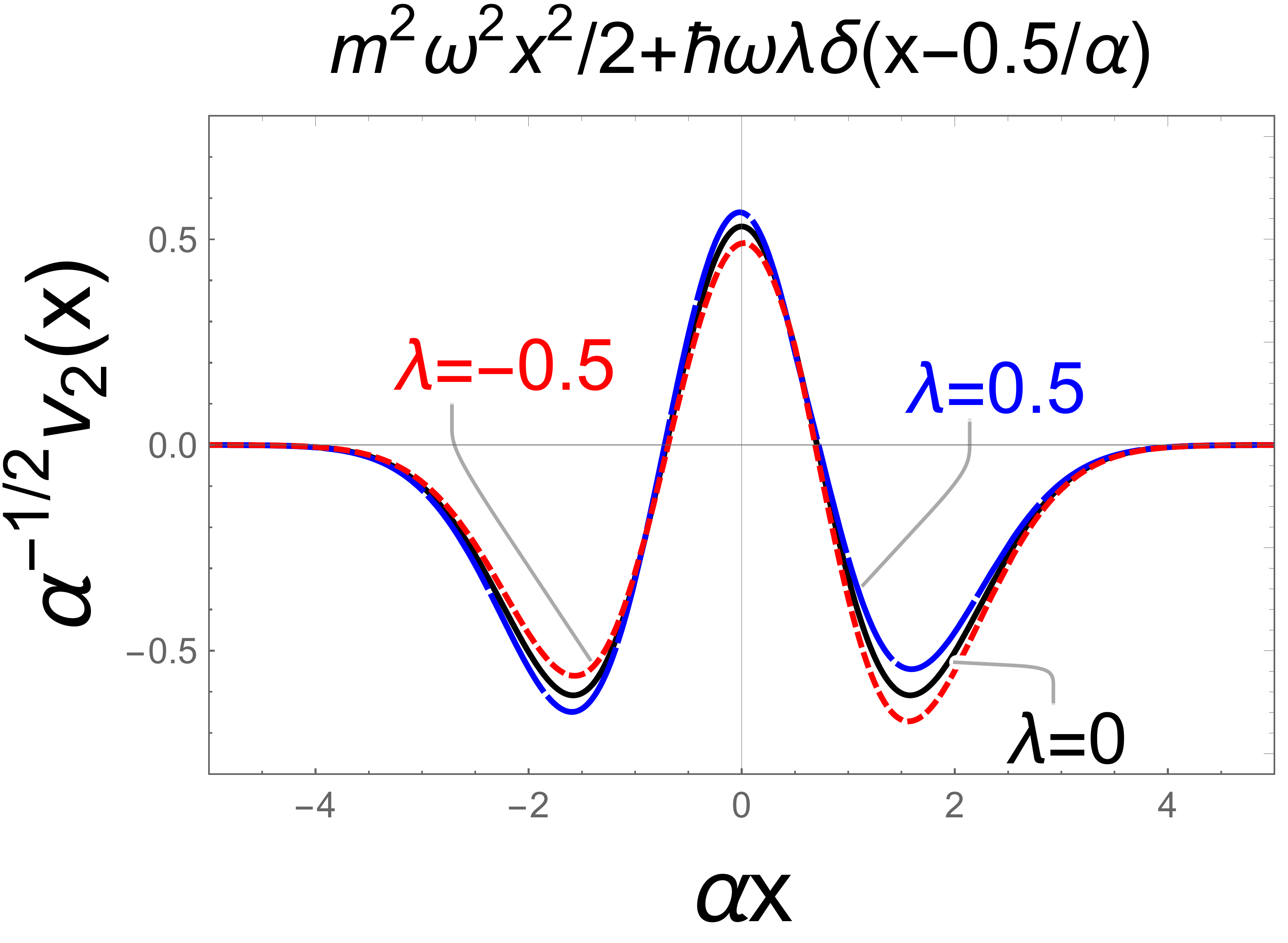}
}
\subfigure[\ $n=3$ 
]{
  \includegraphics[width=6.5cm]{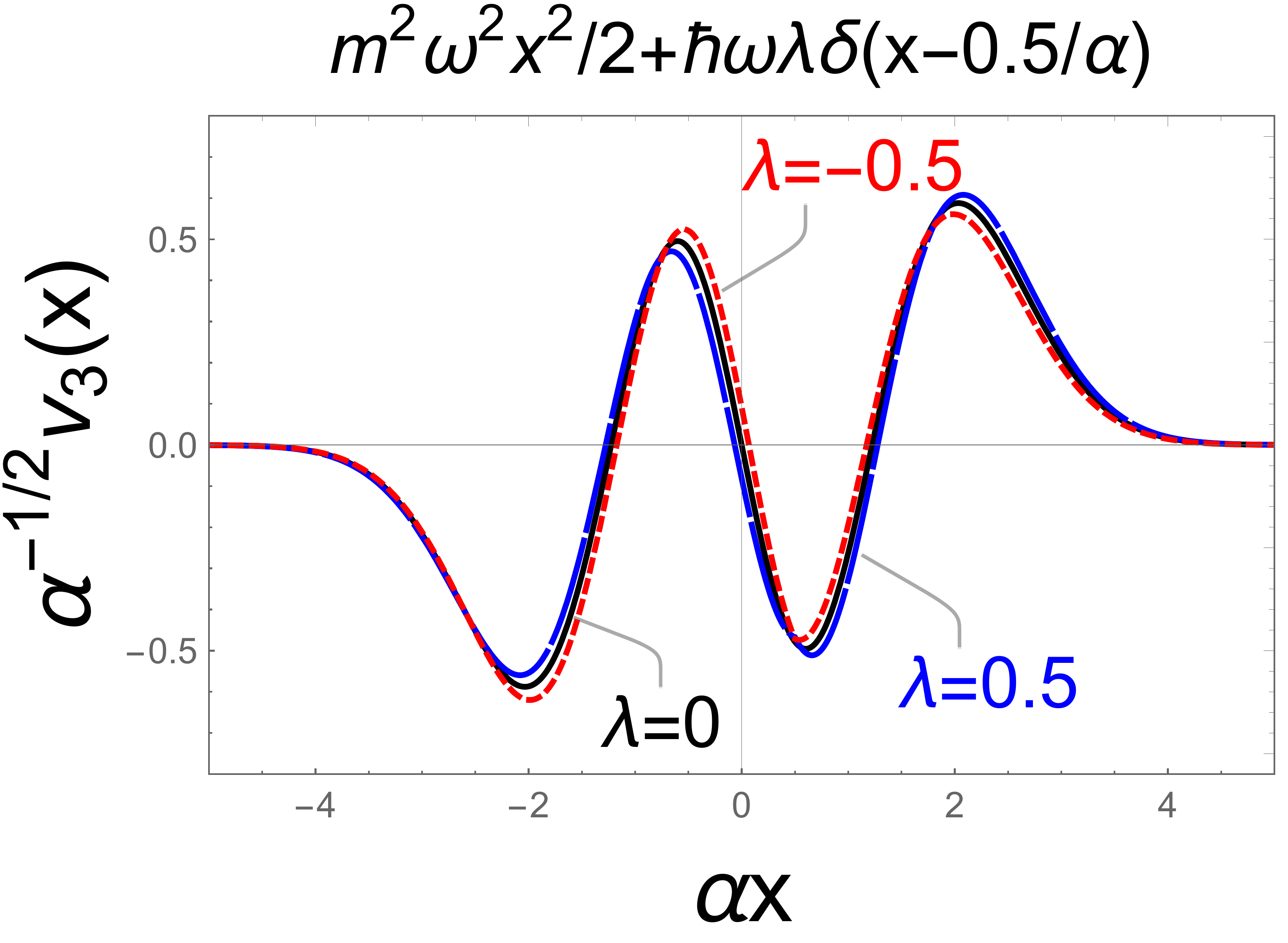}
}
\subfigure[]
{
  \includegraphics[width=9cm]{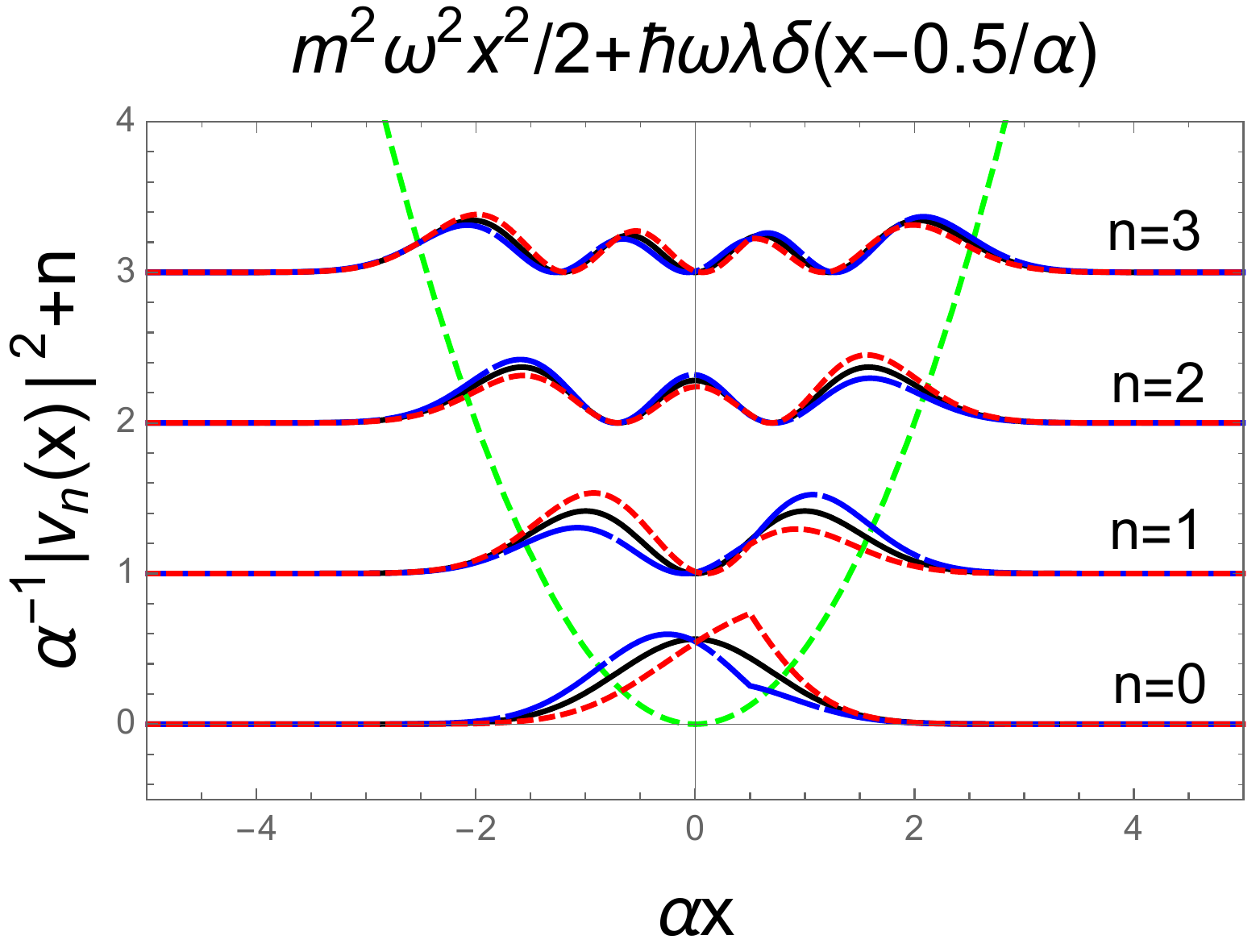}
}
\caption{
  (a)-(d): The plots of 
wave functions $\alpha^{-1/2}v_n(x)$ corresponding to the first four energy levels $n=0,1,2,3$ with $\lambda=0,\pm 0.5$, respectively are shown.
(e): The absolute square values of  
wave functions, $\alpha^{-1}|v_n(x)|^2$, corresponding to the first four energy levels $n=0,1,2,3$ with 
               $\lambda =0.5$ (long-dashed), $\lambda=-0.5$ (dashed), and $\lambda=0$ (solid), respectively are shown. 
   } 
 \label{fig: u0123y1}
\end{figure}

With the help of Eq.~(\ref{eq: WHnu}) we obtain the following transcendental equation from Eq.~(\ref{eq: matchingv}):
\be
\frac{2^{\varepsilon+1/2}\sqrt\pi}{\Gamma(\frac{1}{2}-\varepsilon)}+
2\lambda e^{-\alpha^2a^2}H_{\varepsilon-1/2}(-\alpha a)H_{\varepsilon-1/2}(\alpha a)=0,
\en
which is exactly the same as in Eq.~(\ref{eq: dis1}).
This equation gives us the relation between the coupling strength $\lambda$ and the energy (in $\hbar\omega$ unit) eigenvalue $\varepsilon=\varepsilon_n$, and for a given $\lambda$, we can get the eigenvalue by numerically solving this equation. 

In Table.~\ref{tab:WdeltaEn}, we show the first six 
energy eigenvalues 
$\varepsilon_n-1/2$ for $\alpha a=0.5$ and $\lambda=0,\pm0.5,\pm1$, respectively. 
We 
see that the eigenvalues increase with increasing the coupling strength $\lambda$ and never cross over the adjacent energy levels.

In Fig.~\ref{fig: u0123y1}(a)-(d), we plot the the wave functions $v_n(x)$ corresponding to the first four energy levels $n=0,1,2,3$ with $\lambda=0,\pm 0.5$, respectively.   
It is apparent to see the discontinuity of the first derivative of $v_0(x)$ with $\lambda=\pm 0.5$ at the $\alpha x=0.5$ in Fig.~\ref{fig: u0123y1}(a), but not so apparent in Fig.~\ref{fig: u0123y1}(b-d).
We also see that the discontinuity disappears as the coupling strength $\lambda$ goes to 0. 
On the other hand, $\lambda<0$ ($\lambda >0$) corresponds to attractive (repulsive) force so that we have a larger (smaller) amplitude of wave function at $\alpha x=0.5$ with $\lambda=-0.5$ (0.5) than the original one in Fig.~\ref{fig: u0123y1}(a). 
In Fig.~\ref{fig: u0123y1}(e), we plot the absolute square values of  
wave functions, $|v_n(x)|^2$, corresponding to the first four energy levels $n=0,1,2,3$ with $\lambda =0.5$ (long-dashed), $\lambda=-0.5$ (dashed), and $\lambda=0$ (solid), respectively. We see that there is more probability to appear on the right-hand side of the line $x=0$ than the left-hand for even $n$ and it is reversed for odd $n$ for $\lambda=-0.5$. On the contrary, there is fewer probability to appear on the right-hand side of the line $x=0$ than the left-hand for even $n$ and it is reversed for odd $n$ for $\lambda=0.5$.

\section{Solutions of Schr\"odinger equation of a simple harmonic oscillator system with two generic delta-function potentials}

In this section, we generalize the technics to solving the quanatum system of SHO with two generic delta-function potentials. The corresponding Schr\"odinger equation is
\be
\left(-\frac{\hbar^2}{2m}\frac{\partial^2}{\partial x^2}+V(x) \right)v( (x; y_1,y_2)=E v(x; y_1,y_2),
\en
where
\be
V(x)=\frac{1}{2}m\omega^2 x^2+\frac{\alpha\hbar^2}{m}\lambda_1\delta(x-y_1)+\frac{\alpha\hbar^2}{m}\lambda_2\delta(x-y_2).
\en
Without loss of generality, we have assumed that $y_1<y_2$ in the above.
Similarly, after changing the variable as in Eq.~(\ref{eq: cvariables}), we have
\be
\left(\frac{\partial^2}{\partial x^2}-\alpha^4 x^2+2\alpha^2\varepsilon \right)v(x;y_1,y_2)=2\alpha\lambda_1\delta(x-y_1)v(x;y_1)+2\alpha\lambda_2\delta(x-y_2)v(x;y_2).
\label{eq: SchEq-2}
\en
The wave function needs to satisfy the boundary conditions:
\be
\lim_{x\to\pm\infty}v(x;y_1,y_2)=0,
\en
and the matching conditions around $x=y_1$ and $x=y_2$:
\be
v(x=y_1^+;y_1,y_2)=v(x=y_1^-;y_1,y_2),
\quad
v(x=y_2^+;y_1,y_2)=v(x=y_2^-;y_1,y_2),
\en
\be
\frac{\partial}{\partial x}v(x=y_1^+;y_1,y_2)-\frac{\partial}{\partial x} v(x=y_1^-;y_1,y_2)=2\alpha\lambda v(y_1;y_1,y_2),
\non\\
\frac{\partial}{\partial x}v(x=y_2^+;y_1,y_2)-\frac{\partial}{\partial x} 
v(x=y_2^-;y_1,y_2)=2\alpha\lambda v(y_2;y_1,y_2).
\label{eq: matchingv2}
\en
The wave function is assumed to have the following form:
\be
v(x;y_1,y_2)&=&N f_<\left(\min(x,y_1)\right) f_\parallel(\min(\max(x,y_1),y_2)) f_>\left(\max(x,y_2)\right).
\label{eq: w12}
\en
where $N$ is the normalization constant and 
\be
f_<(x)&=&e^{-\frac{\alpha^2x^2}{2}} H_{\nu}(-\alpha x),
\non\\
f_\parallel(x)&=&e^{-\frac{\alpha^2x^2}{2}} [H_{\nu}(-\alpha x)+\beta H_{\nu}(\alpha x)],
\non\\
f_>(x)&=&e^{-\frac{\alpha^2x^2}{2}} H_{\nu}(\alpha x).
\label{eq: f3}
\en
In the above, $\nu=\varepsilon-1/2$ and the coefficient $\beta$ are undetermined.
The discontinuities of the first derivatives of the wave function $v(x)$ at $x=y_1$ and $x=y_2$ [see Eq.~(\ref{eq: matchingv2})] give us a set of two simultaneous equations:
\be
W\left(f_<(y_1),f_\parallel(y_1)\right)
&=&2\alpha\lambda e^{-\alpha^2y_1^2}H_{\varepsilon-1/2}(-\alpha y_1)H_{\varepsilon-1/2}(\alpha y_1),
\non\\
W\left(f_\parallel(y_2),f_>(y_2)\right)
&=&2\alpha\lambda e^{-\alpha^2y_2^2}H_{\varepsilon-1/2}(-\alpha y_2)H_{\varepsilon-1/2}(\alpha y_2).
\en
For a given coupling strength $\lambda$, we can numerically solve the eigenvalue $\varepsilon=\varepsilon_n$ and 
the coefficient $\beta$ from the above two simultaneous equations.
For simplicity, we consider the case: $\lambda_1=\lambda_2\equiv\lambda$ and $y_1=-y_2$ so that the potential is symmetric, i.e. $V(-x)=V(x)$, and hence the wave function $v_n(x)$ must be even or odd. From Eq.~(\ref{eq: w12}) and 
Eq.~(\ref{eq: f3}), we know that $\beta=1$ for even function $v_n(x)$ and $\beta=-1$ for the odd function $v_n(x)$. 
We note that in passing the above transcendental equations are much simpler than those shown in ref.~\cite{FB}.
\begin{table}[t!]
\centering
\captionsetup{justification=raggedright}
{\begin{tabular}{|c|c|c|c|c|c|c|}
\hline
$\varepsilon_n-1/2$
         & $n=0$ & $n=1$ & $n=2$ & $n=3$ & $n=4$ & $n=5$
          \\  \hline
$\lambda=0$
         & 0 & 1 & 2 & 3 & 4 & 5
          \\  \hline
$\lambda=-0.5$
         &-0.49476& 0.711225& 1.9511& 2.75301& 3.99888& 4.81243
          \\  \hline 
$\lambda=0.5$          
          & 0.389598& 1.17256& 2.06173& 3.2031 & 4.00117& 5.18881
           \\  \hline 
$\lambda=-1$  
          &-1.11286& 0.230993& 1.91252 & 2.50163& 3.9978& 4.64815
           \\  \hline
$\lambda=1$  
          & 0.689733 & 1.28047& 2.13806& 3.35457 & 4.00238 & 5.35735
           \\  \hline                                                                                                                                                  
\end{tabular}
}
\caption{The first six  
energy eigenvalues $\varepsilon_n-1/2$ for $\alpha y_1=0.5, \alpha y_2=-0.5$ and $\lambda_1=\lambda_2\equiv\lambda=0,\mp 0.5,\mp 1$ are shown.}
\label{tab:WdeltaEn2}
\end{table}
\begin{figure}[t!]
\centering
\captionsetup{justification=raggedright}
\subfigure[\ $n=0$\ 
]{
  \includegraphics[width=6.5cm]{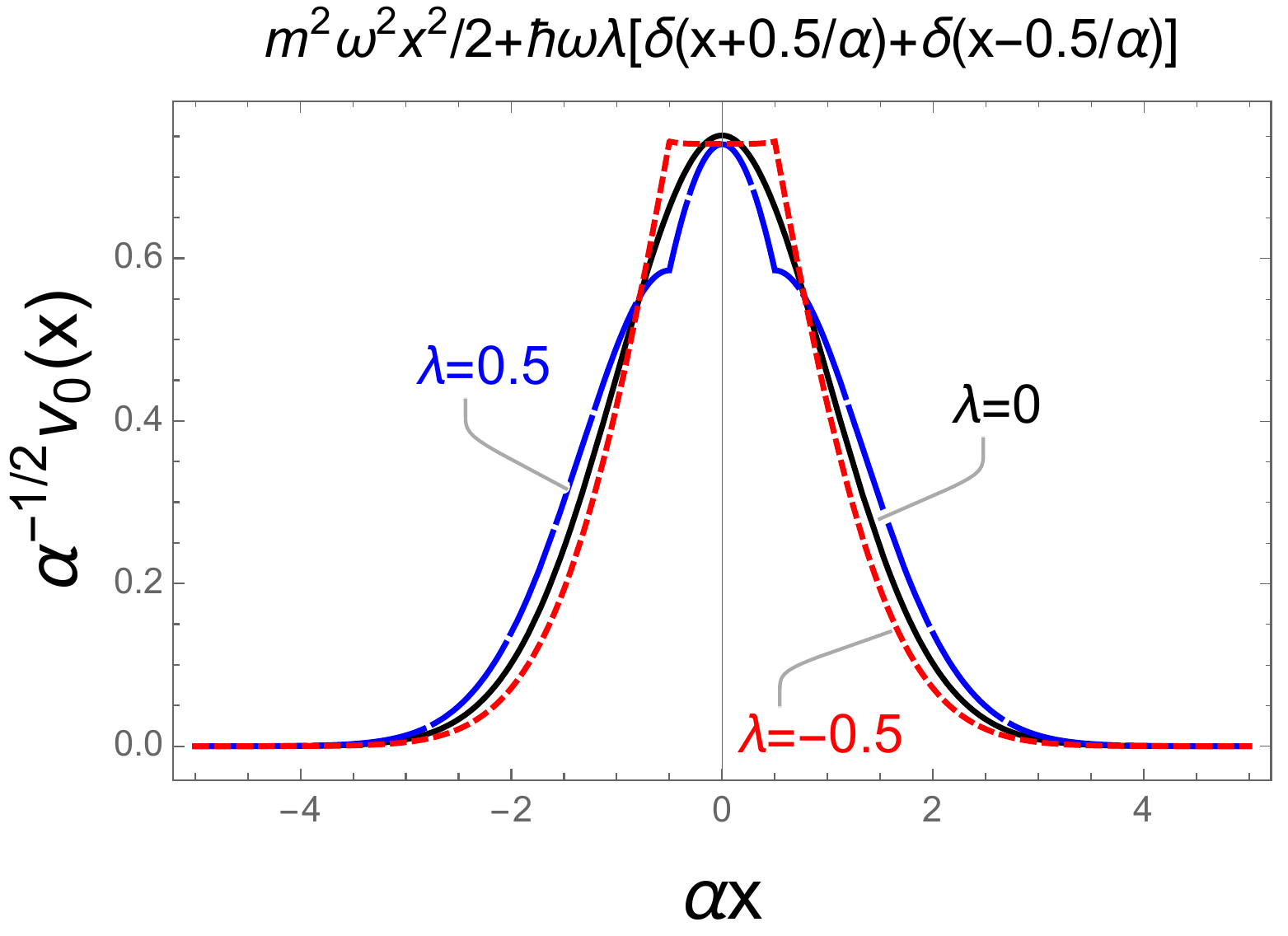}
}
\subfigure[\ $n=1$\ 
]{
  \includegraphics[width=6.5cm]{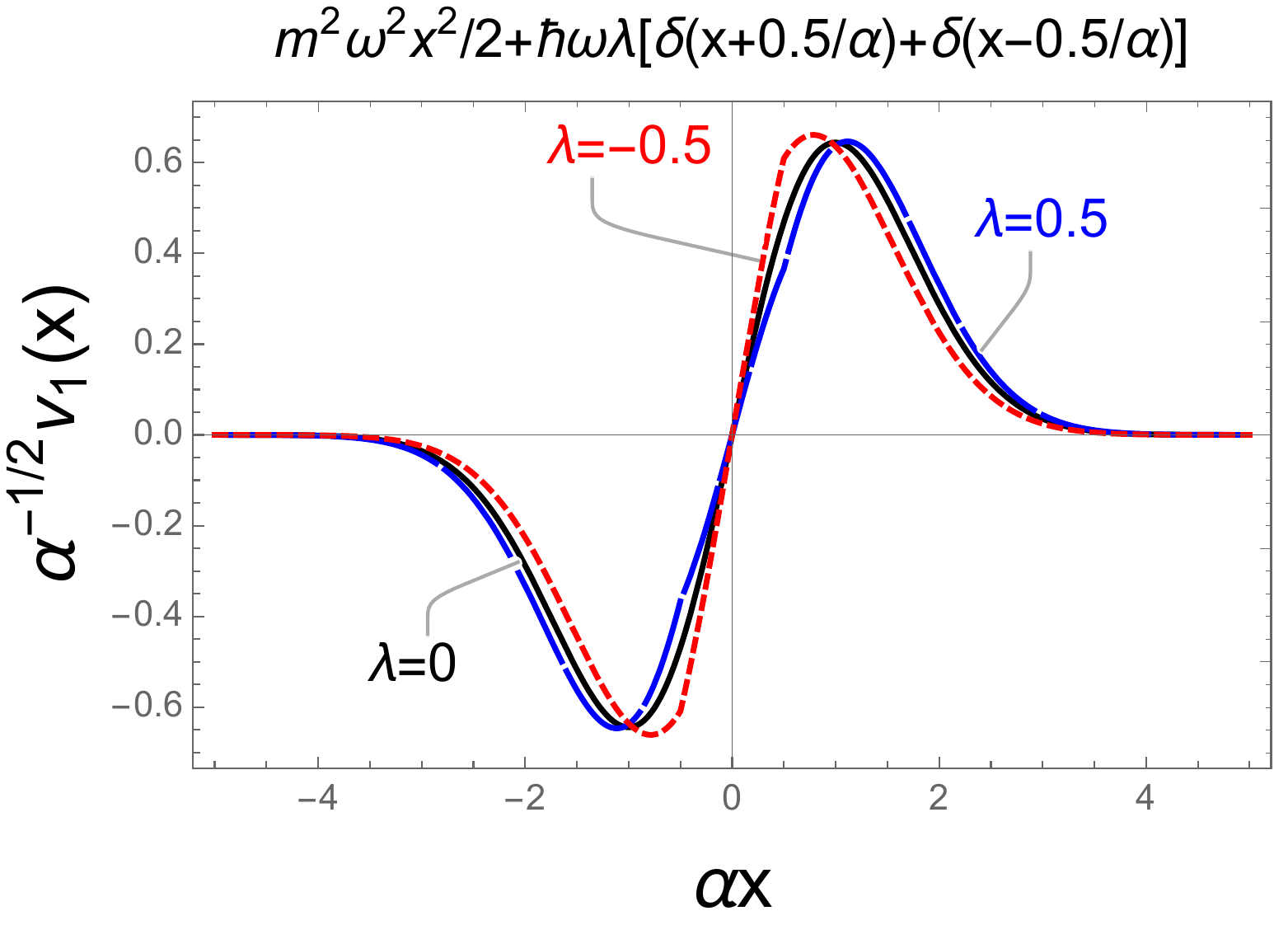}
}
\subfigure[\ $n=2$\
]{
  \includegraphics[width=6.5cm]{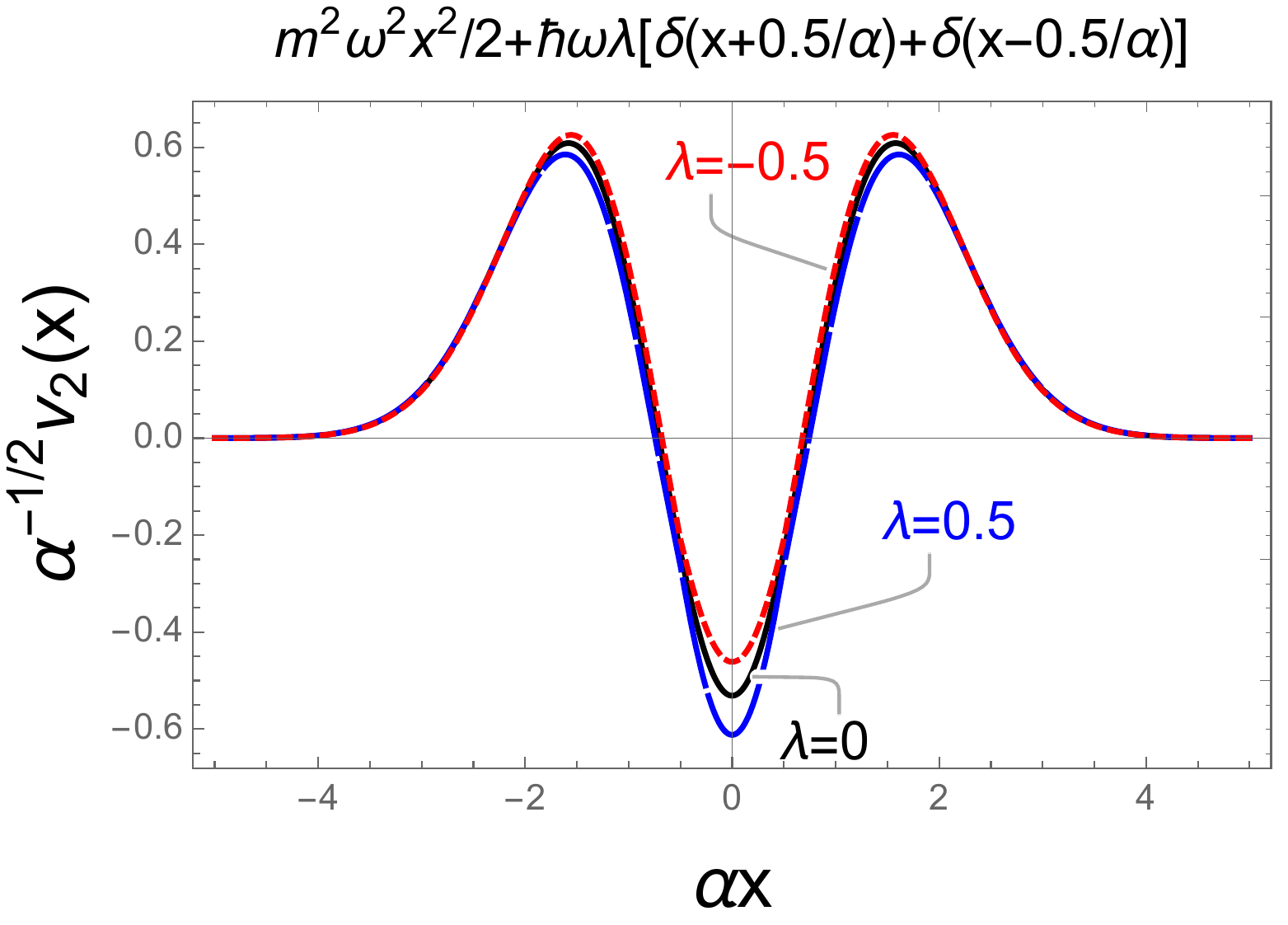}
}
\subfigure[\ $n=3$\ 
]{
  \includegraphics[width=6.5cm]{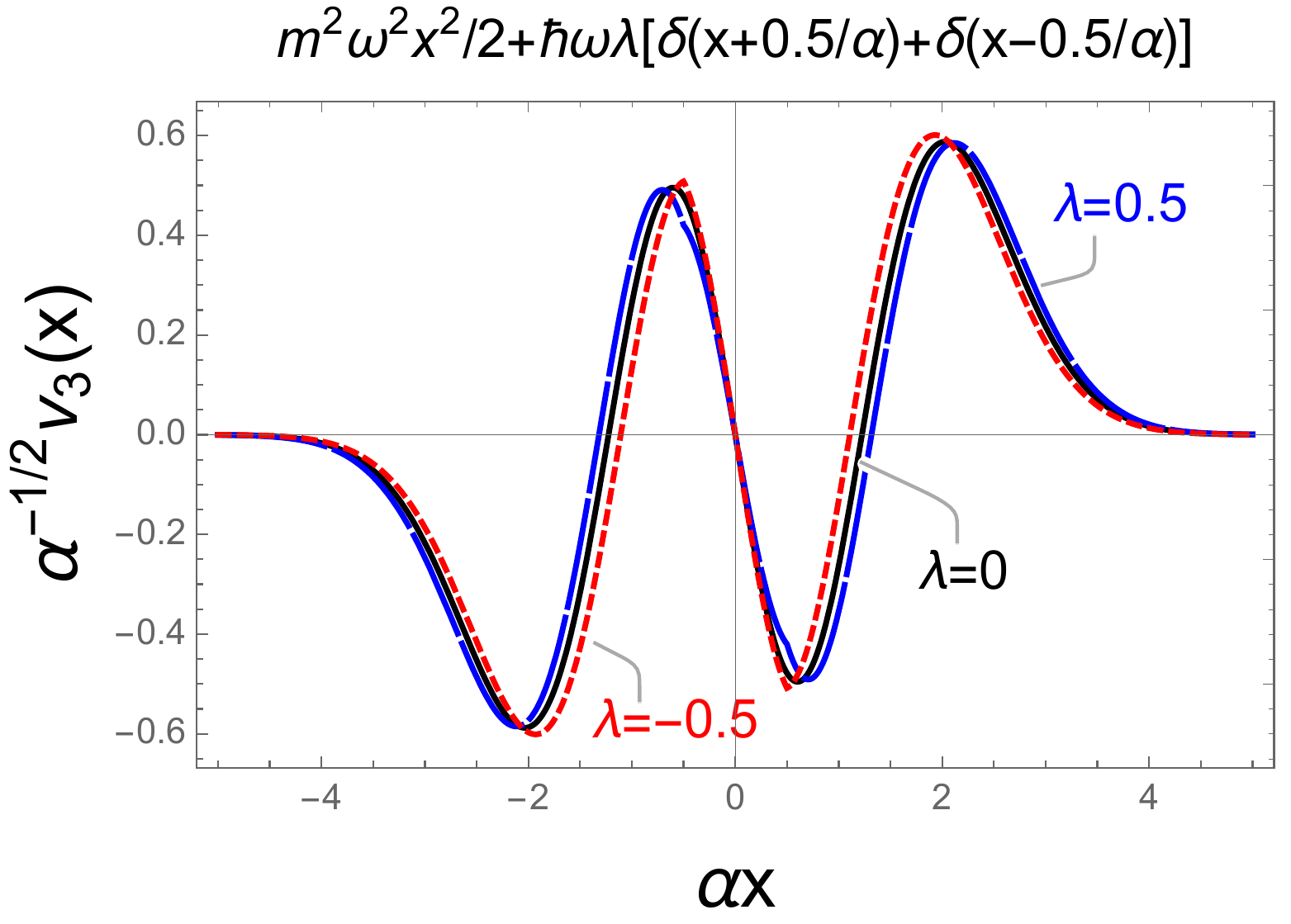}
}
\subfigure[]{
  \includegraphics[width=9cm]{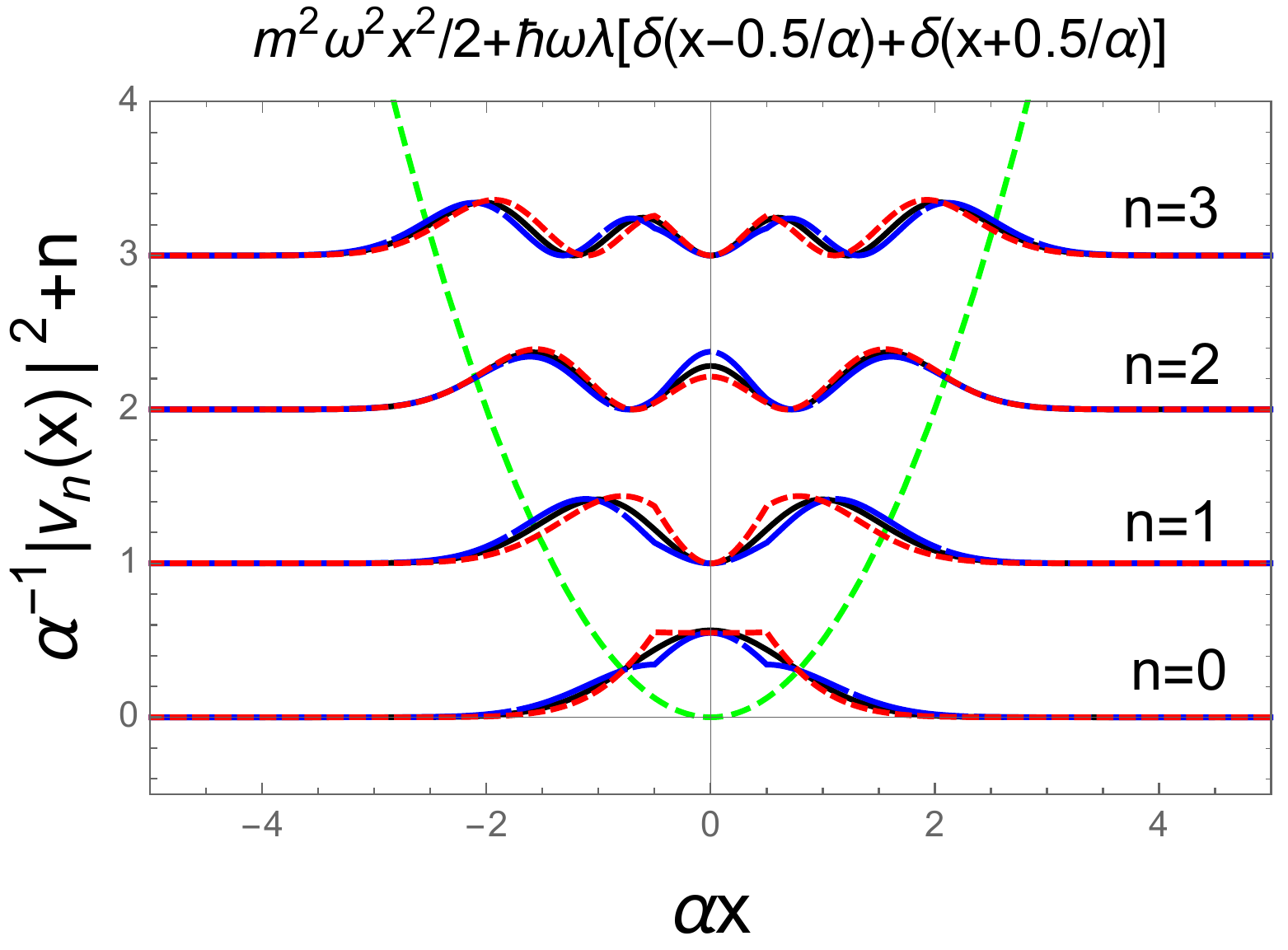}
}
\caption{
 (a)-(d): The plots of wave functions $\alpha^{-1/2}v_n(x)$ corresponding to the first four energy level $n=0,1,2,3$ with $\lambda=0,\pm 0.5$, respectively are shown. 
(e): The absolute square values of wave functions, $\alpha^{-1}|v_n(x)|^2$, corresponding to the first four energy levels $n=0,1,2,3$ with $\lambda =0.5$ (long-dashed), 
$\lambda=-0.5$ (dashed), and $\lambda=0$ (solid), respectively are shown.
   } 
 \label{fig: u0123y2}
\end{figure}

In Table.~\ref{tab:WdeltaEn2}, we show the first six energy eigenvalues 
$\varepsilon_n-1/2$ for $\alpha y_1=-0.5,\ \alpha y_2=0.5$ and $\lambda=0,\pm0.5,\pm1$, respectively. 
We 
also see that the eigenvalues increase with increasing the coupling strength $\lambda$ and never cross over the adjacent energy level as in the previously case.
In Fig.~\ref{fig: u0123y2}(a)-(d), we plot the the wave functions $v_n(x)$ corresponding to the first four energy levels $n=0,1,2,3$ with $\lambda=0,\pm 0.5$, respectively. 
Since the Hamiltonian is parity invariant, the wave function $v_n(x)$ is either even or odd.
It is apparent to see the discontinuity of the first derivatives of the wave function $v_n(x)$ at the $\alpha x=\pm 0.5$ with $\lambda=\pm 0.5$ in Fig.~\ref{fig: u0123y2}(a), but not so apparent in Fig.~\ref{fig: u0123y2}(b-d). 
We also see that the discontinuity disappears as the coupling strength $\lambda$ goes to 0.  
In Fig.~\ref{fig: u0123y2}(e), we plot the absolute square values of 
wave functions, $|v_n(x)|^2$, corresponding to the first four energy levels $n=0,1,2,3$ with $\lambda =0.5$ (long-dashed), $\lambda=-0.5$ (dashed), and $\lambda=0$ (solid), respectively.
Since $\lambda<0$ ($\lambda >0$) corresponds to the attractive (repulsive) force, in general, we have a larger (smaller) square amplitude of wave function at $\alpha x=\pm0.5$ than the $\lambda=0$ case.

\section{Conclusions}

We study the one-dimensional problem on the quantum system of SHO without/with one (or two) generic delta-function potential(s).
For the pure quantum SHO system, we derive a complete analytical form for the corresponding time-independent Green's function. 
It is natural and straightforward to obtain eigenvalues and eigenfunctions from the Green's function. 
The energy eigenvalues can be obtained from the poles of the Green's function and using the residue theorem, we can obtain the familiar SHO wave functions. 
We see that the Wronskian plays the interesting and important roles in $G_0$ and the wave function $u_\nu$ [see Eq.~(\ref{eq: unu})]. 
Although it is possible to obtain $G_0(x,y; E)$ by performing the Fourier transform of 
$K_0(x,t;y,t_0)$ using an integral of Bessel function~\cite{Kleinert:2004ev}, 
the above interesting point will be obscured.
For the quantum SHO system with a generic delta-function potential, we follow the approach of ref.~\cite{Cavalcanti} to obtain the time-independent Green's function from the SHO Green's function obtained previously. 
The eigenvalues can be obtained from the poles of the Green's function. 
Our results agree with \cite{{Viana-Gomes}}, but our method can also be applied to a delta-function potential at an arbitrary site.
Nevertheless, the simplest way to find the wave functions is to solve the Schr\"odinger equation directly. 
In fact, the same technics of solving the Green's function $G_0$ can be used to solve the eigenvalue problem of the simple harmonic oscillator with an generic delta-function potential at an arbitrary site. 
The technics can be easily generalized to solve the quantum system of SHO with two generic delta-function potentials.
For illustration, we solve the case of symmetric potential and obtain energy eigenvalues and eigenstates for the first few states. 

\section*{Acknowledgments}

The authors are grateful to Chuan-Tsung Chan and Kwei-Chou Yang for discussions.
This research was supported by the Ministry of Science and Technology of R.O.C. under Grant Nos. 105-2811-M-033-007 and in part by 103-2112-M-033-002-MY3 and 106-2112-M-033-004-MY3.

\end{document}